\documentclass[fleqn,usenatbib]{mnras}

\usepackage[T1]{fontenc}

\DeclareRobustCommand{\VAN}[3]{#2}
\let\VANthebibliography\thebibliography
\def\thebibliography{\DeclareRobustCommand{\VAN}[3]{##3}\VANthebibliography}

\usepackage{graphicx}	% Including figure files
\usepackage{amsmath}	% Advanced ma/ths commands
\usepackage{amssymb}	% Extra maths symbols
\usepackage{orcidlink}
\usepackage{txfonts}
\usepackage{listings}
\usepackage{nameref}
\usepackage{rotating}
\usepackage[utf8]{inputenc}
\usepackage{hyperref} %\usepackage[draft]{hyperref}
\usepackage{booktabs}
\usepackage{xcolor}
\usepackage{adjustbox}
\usepackage{xspace}
\usepackage{academicons}

\usepackage{newtxtext,newtxmath}
\definecolor{rev}{HTML}{1F456E}

\definecolor{mygreen}{rgb}{0,0.6,0}
\definecolor{mygray}{rgb}{0.5,0.5,0.5}
\definecolor{mymauve}{rgb}{0.58,0,0.82}
\title[NIRDust]{NIRDust: Probing Hot Dust Emission around Type 2 AGN using K-band Spectra}
\author[G. Gaspar et al.]{
Gaia Gaspar $^{1, 5}$ \orcidlink{0000-0001-9293-4449}\thanks{E-mail: ggaspar@unc.edu.ar}
Martín Chalela $^{1, 2, 5}$ \orcidlink{0000-0002-9143-7974}
Juan Cabral $^{2, 3, 5}$ \orcidlink{0000-0002-7351-0680}
José Alacoria $^{4}$ \orcidlink{0000-0001-9726-8991}
\newauthor
Damián Mast $^{1, 5}$ \orcidlink{0000-0003-0469-3193}
and Rubén Díaz $^{1, 6}$ \orcidlink{0000-0001-9716-5335}
\\
$^{1}$Observatorio Astronómico de Córdoba, Universidad Nacional de Córdoba, Laprida 854, X5000BGR, Córdoba, Argentina (OAC--UNC).\\
$^{2}$Instituto De Astronom\'ia Te\'orica y Experimental,  C\'ordoba, Argentina (IATE--CONICET).\\
$^{3}$Gerencia De Vinculaci\'on Tecnol\'ogica Comisi\'on Nacional de Actividades Espaciales (CONAE), Falda del Ca\~nete, C\'ordoba, Argentina.\\
$^{4}$Instituto de Ciencias Astr\'omicas, de la Tierra y el Espacio, Av. España Sur 1512, San Juan, Argentina (ICATE--CONICET--UNSJ).\\
$^{5}$Consejo de Investigaciones Cient\'{i}ficas y T\'ecnicas de la Rep\'ublica Argentina, Buenos Aires, Argentina (CONICET).\\
$^{6}$Gemini Observatory, NSF's NOIRLab, USA.
}

\date{Accepted 21 Dec 2023}

\pubyear{2023}

\begin{document}
\label{firstpage}
\pagerange{\pageref{firstpage}--\pageref{lastpage}}
\maketitle

% juanchito achico el texto por que nos pasabamos por mas de 10 palabras
% tienen que ser 250 max
\begin{abstract}
Hot dust in the proximity of AGNs strongly emits in the Near Infrared producing a red excess that, in type 2 sources, can be modeled to measure its temperature. In the era of high spatial-resolution multi-wavelength data, mapping the hot dust around Supermassive Black Holes is important for the efforts to achieve a complete picture of the dust role and distribution around these compact objects.
In this work we propose a methodology to detect the hot dust emission in the proximity of Type 2 AGNs and measure its temperature using K-band spectra ($\lambda_c$ = 2.2\,$\mu$m).
To achieve this, we have developed NIRDust, a Python package for modeling K-band spectra, estimate the dust temperature and characterize the involved uncertainties. We tested synthetic and real spectra in order to check the performance and suitability of the physical model over different types of data.
Our tests on synthetic spectra demonstrated that the obtained results are influenced by the signal-to-noise ratio (S/N) of the input spectra. However, we accurately characterized the uncertainties, which remained below $\sim$150\,K for an average S/N per pixel exceeding 20. Applying NIRDust to NGC\,5128 (Centaurus A), observed with the Gemini South Telescope, we estimated a dust temperature of 662 and 667\,K from Flamingos-2 spectra and 697 and 607 \,K from GNIRS spectra using two different approaches.

\end{abstract}

% se supone que A&A usa las mismas keywords que mnras
\begin{keywords}
ISM: dust --
Infrared: ISM --
Galaxies: nuclei --
Galaxies: active --
Techniques: spectroscopic --
Methods: data analysis --
\end{keywords}

%%%%%%%%%%%%%%%%%%%%%%%%%%%%%%%%%%%%%%%%%%%%%%%%%%

%%%%%%%%%%%%%%%%% BODY OF PAPER %%%%%%%%%%%%%%%%%%

% =============================================================================
% Section Intro
% =============================================================================

\section{Introduction}
\label{sec:intro}

The Unified Model for Active Galactic Nuclei \citep{Antonucci1993, Netzer2015} proposes orientation with respect to the observer as the fundamental difference between AGN types. The key to the unification is a nuclear structure of molecular gas and dust that absorbs the radiation in some lines of sight but not in others and was unresolved in the observations that drove the model \citep[e.g.][]{Mason2015}.  As modern facilities become available this structure can be studied at higher spatial resolutions, via infrared spectroscopy and interferometry and sub-mm observations, bringing clues to its true morphology. This structure, called "the torus" due to its theoretical origin, not only absorbs radiation in order to display the differences between AGN types but it could play an important role in the accretion mechanism of the SMBH as it is part of the inflow and outflow processes in the nuclear regions of galaxies. 

Over the last decades the model for the dusty absorber has suffered profound changes. The idea of a “clumpy torus” was adopted after the homogeneity of the torus was ruled out on the basis of the high temperatures the dust reaches under such conditions \citep{krolik1988, nenkova2002, nenkova2008}. Later, the discovery of a polar component observed with MIR interferometry \citep[e.g.][]{Honig2012, Honig2013, LopezGonzaga2014, Asmus2016, Stalevski2019,Asmus2019} resulted in a new model that incorporates a polar component as an outflowing wind and a thin annulus of high temperature dust ($\sim$\,1500 K) observed in NIR interferometry data at the dust sublimation radius \citep{Honig2019}. Parallel to this, several authors have found massive and large equatorial components observing in the sub-mm regime, establishing that the dusty absorber can reach sizes up to 100 pc \citep[e.g.][]{Izumi2018, Alonso-Herrero2018, combes2019, Garcia-burillo2021}. Regarding the chemical composition of the dust, temperature plays a key role, as different grain composition and sizes sublimate with increasing temperature. For instance, silicates are destroyed in regions with temperatures above 1200\,K while graphite grains can survive up to 1900\,K \citep{Honig2017}.In this context, the temperature of the hot dust in the torus is a fundamental parameter that is related with the determination of the density, the composition and the geometry of the torus.

Dust radiates at different wavelengths depending on its temperature. In the case of dust thermally heated by the accretion disc of a SMBH part of the emission is detected in the near-infrared and, in particular, is prominent in the  K-band (2-2.4$\mu$m). Several authors have reported measurements of the temperature of hot dust in AGNs using JHK colors \citep[e.g.][]{Glass1985,AlonosoHerrero1998}, NIR interferometry \citep[e.g.][]{Kishimoto2011,Leftley2021, Gravity2020, Gamez-Rosas2022} or NIR spectroscopy \citep[e.g.][]{Burtscher2015, Durre2018, Gaspar2019, Gaspar2022, Riffel2022} and found temperatures ranging from 700 to 1500 K. \citet{Burtscher2015} measured the temperature of the hot dust component for 51 near AGNs in 1” apertures and found a mean of 887\,$\pm$\,68\,K for type 2 AGNs and 1292\,$\pm$\,46\,K for type 1 AGNs. Regarding the extension and temperature distribution of this hot dust, \citet{Gaspar2019, Gaspar2022} found T\,$\sim 1000$\,K dust in scales of tens of pc and almost constant in temperature  in two Seyfert 2 nuclei. \citet{Gamez-Rosas2022} measured dust temperatures using MIR interferometry with MATISSE/ESO/VLTI instrument in several sub-pc regions around the SMBH of NGC\,1068, the archetypical Seyfert 2 galaxy. They found temperatures ranging between 800 and 1500\,K in the different regions that “do not decrease systematically with distance from the inferred black hole position”. Finally, \citet{Riffel2022} found hot dust emission extended outside the unresolved core and stated that the emission inside $\sim$\,40\,pc is dominated by hot dust in a sample of 18 nearby Seyfert galaxies.
The variety of these dust detections evidence that more temperature measurements in high angular resolution infrared data are necessary to improve our understanding on the extension, distribution, and origin of the hot dust in the proximities of AGNs. 

The resolved NIR spectroscopic observations currently available with ground-based 8-10\,m class telescopes are suitable for determining the extent and temperature distribution of the hot dust component in AGNs, as they can resolve the central regions of these objects at resolutions of tens of pc or less in nearby galaxies. Furthermore, the MIRI (5-28 $\mu$m) and NIRSpec (0.6 - 5 $\mu$m) instruments on board the James Webb Space Telescope will enable spectroscopic mapping of dust and molecular gas in the central regions of AGNs with unprecedented sensitivity and spatial resolution.

This work introduces NIRDust, a Python-based tool designed for detecting and measuring the temperature of hot dust near Type 2 AGNs. NIRDust utilizes K-band rest frame spectra to fit the temperature of a blackbody component emitted by the hot dust in the NIR range. The package offers a range of functionalities, including spectrum storage, pre-processing, and fitting. NIRDust's application programming interface (API) facilitates seamless integration with Astropy \citep{Astropy}, particularly with Specutils, enabling users to incorporate NIRDust into their broader code base when working with NIR spectra.

In Section~\ref{sec:physics} we present the fundamental physics basis for the code. In Section~\ref{sec:tech} the functionalities  and quality of the package are described along with an application example and, in Section~\ref{sec:performance} the results of tests to evaluate the performance of NIRDust are presented. The Summary and future work are in Section~\ref{sec:conclusions}.

% =============================================================================
% Physics
% =============================================================================
\section{Modelling the hot dust emission}
\label{sec:physics}

For type 2 AGNs, in the absence of a non-thermal featureless continuum, the nuclear spectrum continuum in the K-band is conformed by two elements: the stellar population spectrum and the emission by hot dust \citep[][and references therein]{Thatte1997, Gratadour2003}. This is easily appreciated when observing the slope of the nuclear spectrum continuum that remains constant or even rising when a pure stellar spectrum is expected to fall \citep[e.g.][]{Ferruit2004, Gravity2020}, i.e. a red excess is present. This means that, if the stellar spectrum can be computed then the dust emission can be isolated. This hot dust component is not only detected in the nuclear spectrum but it can be found in more external regions in high-angular resolution spectra \citep[e.g.][]{Gaspar2022, Riffel2022}.

At moderate spectral resolution ($R=1000-3000$), the dust emission can be modeled by a blackbody function of single temperature. In the real case, the emission could be composed by a continuum of temperatures, but for this kind of spectra experience has shown that using a single temperature is sufficient \citep[e.g.][]{{Durre2018},{Gaspar2019}}.

The blackbody radiation can be described by Planck's law as a function of frequency $\nu$, and parameterized by its temperature, $T$, and an amplitude, $A$, as:
\begin{equation}
    B_T(\nu) = A \frac{2h\nu^3}{c^2}\frac{1}{e^{h\nu/kT} - 1}
\end{equation}
where $h$, $k$ and $c$ correspond to the Planck constant, Boltzmann constant and speed of light, respectively. The scale factor $A$ is used to model the intensity amplitude of the blackbody radiation for a given temperature $T$. It is worth noticing that the introduction of this factor does not affect the location of the radiation peak or the general shape of the blackbody function. The dust emission can then be modeled by the free parameters $A$ and $T$.

To model the Target Spectrum, where the hot dust emission is expected to be found, we use a linear combination of the stellar component and a blackbody function as in \citet{Durre2018}:

\begin{equation}
\label{eq2}
\begin{aligned}
    S_t & = \alpha S_s + \beta^{\prime} B_T + \gamma^{\prime} \\
    \beta^{\prime} & = 10^{\, \beta} \\
    \gamma^{\prime} & = 10^{\, \gamma}
\end{aligned}
\end{equation}

Here, $S_t$ is the Target Spectrum, $S_s$ is the Reference Spectrum that accounts for the stellar emission, $B_T$ is the blackbody function of temperature T, and $\gamma^{\prime}$ is a constant emission that represents the sky background (it is recommended that the sky background is previously subtracted to make $\gamma^{\prime}$ the closer to zero as possible). $\alpha$ and $\beta^{\prime}$ are scale factors of the stellar and blackbody components respectively. The reformulation of $\beta^{\prime}$ and $\gamma^{\prime}$ is introduced for the technical purpose of exploring the parameter space with a logarithmic distribution (see Section \ref{sec:tech} for more details on the technical discussion). Thus, the parameter space is defined by $\boldsymbol{\theta} \equiv \{T, \alpha, \beta, \gamma\}$.

In order to model the hot dust emission with Eq. \ref{eq2} spectra do not necessarily have to be flux calibrated. This is important because the uncertainties introduced in the flux-calibration process can reach $30-50\%$ in the NIR due to the fast changes in atmospheric conditions that occur at this wavebands. The telluric correction (essential when studying spectral continuum) eliminates the shape of the atmosphere transmission and the instrumental response from the spectrum and hence the flux calibration is reduced to a multiplicative operation. Given the linear combination that models the Target Spectrum, this multiplicative factor is absorbed by the $\beta$ parameter. The temperature information does not change after the multiplicative operation.

\subsection{The Reference Spectrum}
\label{subsec:reference spectrum}
As stated above, in order to isolate the hot dust spectrum present in the nuclear region of a Type 2 AGN, it is necessary to provide the stellar spectrum of the same region. The hypothesis working here is that $S_s$ is the same as the Target Spectrum stellar component, in other words, that the stellar population that emitted $S_s$ is the same stellar population which inhabits the region that emitted $S_t$.

$S_s$ is , of course, not possible to isolate from the Target Spectrum without knowing the dust emission, but there are some hypothesis under which this stellar spectrum can be represented by different proxies. The main hypothesis is that in the NIR the late-type population emission is dominant in comparison with the emitted by young stars. This allows, for instance, for the nuclear stellar population to be represented by a spectrum or a mix of spectra of late-type stars. Moreover, the late-type population is more homogeneously distributed than the young population that is usually located in clusters or "clumps" as can be seen, for example, in Fig.~1 of \cite{Lin2013} were the same galaxy is shown in different wavebands. Taking this into account it could be considered that a spectrum of the stellar population extracted at a certain distance from the nucleus can serve as reference for the nuclear stellar population. 

One crucial consideration of this approximation is the frequent occurrence of starburst activity intertwined with AGN emissions in galactic nuclei. Depending on the intensity and duration of the starburst, deviations from the external stellar population used as a reference can emerge. In this context, the choice of width and location for extracting the Reference Spectrum becomes pivotal in accurately emulating the nuclear stellar population. Notably, in cases where high spatial resolution data is available, analyzing the slopes of the continuum at various radii can offer insights into where the stellar population is most representative of the overall population and where regions of hot dust are prevalent. If a multitude of spectra are extracted from distinct radii, a meticulous examination of the spectral slopes can reveal a radius at which the slope consistently remains unchanged as one moves to larger radii. Once this phenomenon is observed, it becomes reasonable to infer that these particular spectra are indeed representative of the underlying spectral population. This approach provides a valuable tool for pinpointing radii where the stellar population's characteristics remain relatively stable and unaffected by extreme local conditions such as the presence of hot dust. An illustrative example of this concept can be found in Fig. 9 of \citet{Gaspar2022}, where the variations in the slope of the spectral continuum are evident between the nucleus and two distinct radii. Interestingly, in these two radii, the spectral slope remains consistent. For a comprehensive set of spectra illustrating these findings, the interested reader is directed to the appendix of the aforementioned publication. In Sec. \ref{subsec:comparison} of this work, we elaborate on the procedure employed to select the Reference Spectrum for a specific example presented herein, corresponding to the galaxy NGC 5128.

In summary and under the aforementioned hypothesis, $S_s$ could be represented by a Reference Spectrum in three different ways:

\begin{itemize}
    \item A spectrum of the same galaxy extracted in the nuclear region but far enough from the nucleus to avoid the hot dust component. (moderate to high spatial resolution data)
    \item A spectrum of a late-type star.
    \item A nuclear stellar template of a non active galaxy of similar type to the host galaxy under study.
\end{itemize}

It is important to acknowledge that NIRDust has been exclusively tested using the first and second scenarios. The first one, which involves analyzing a spectrum extracted near the nucleus of the same galaxy, has been previously employed in other studies \citep[e.g.][]{Durre2018, Gaspar2019, Gaspar2022} and has also been successfully implemented as an alternative in situations where the Population Synthesis technique does not yield satisfactory results \citep{Dumont2020}. The second scenario, that was applied by \citet{Burtscher2015}, and the third scenario, which remains hypothetical, require further testing to determine their suitability for modeling the stellar population of AGNs using NIRDust.

% =============================================================================
% Technical
% =============================================================================
\section{NIRDust}
\label{sec:tech}

The entire experiment of this work are supported by NIRDust. NIRDust is an object-oriented package \citep{ram2003dr} that is fundamentally designed around the factory-method pattern \citep{gamma1995elements}, so the easiest way to create the available objects on the package, is usually some function or method.
The core features of the project are centered around the \texttt{NirdustSpectrum} class, serving as a comprehensive model of a spectrum. This class is capable of segmenting the spectrum within designated boundaries, transforming the spectral axis into frequencies, standardizing intensity levels, deriving noise from a region defined by the user, and implementing selective masking on distinct portions of the spectrum.
In addition, NIRDust offers additional preprocessing capabilities for line spectrum construction and resampling of two spectra to match the same resolution, 
Finally, the main functionality of the package is also available, which consists in estimating and characterizing the dust temperature, this algorithm is explained in detail in the Subsection~\ref{subsection:fit}.

To guarantee accurate results, NIRDust maintains a strict adherence to the PEP-8 coding standard \citep{van2001pep} and employs the Flake-8 tool for seamless code consistency. With a comprehensive suite of 93 unit tests \citep{Jazayeri2007}, it meticulously validates software components, accommodating Python versions spanning from 3.8 to 3.11, and achieving an impressive 99\% code coverage \citep{miller1963systematic}.

A detailed description of the use of NIRDust, the installation process, the bug report procedure, and a more in-depth discussion of the project quality can be found in Appendix~\ref{Apendix:application_example}.

\subsection{Fitting procedure}
\label{subsection:fit}

The model given by equation \ref{eq2} is fitted using both spectra, Target and Reference, using a basin-hopping algorithm \citep{Basinhopping1997}. This is a two-phase method that combines a global minimization algorithm with local minimization at each step. When the local optimization routine finds a local minimum, the basin-hopping algorithm randomly perturbs this local solution and attempts a new local optimization. This technique has provided useful results in many problems where the likelihood surface is hard to explore and simpler minimization algorithms tend to get stuck in local minima.

The parameter space of the problem, is given by four free parameters $\boldsymbol{\theta} \equiv \{T, \alpha, \beta, \gamma\}$. We consider the case where observations, $D$, are independent and identically distributed, sampled from the same gaussian probability function $f(D | \boldsymbol{\theta})$, where each parameter is described by a normal distribution $\mathcal{N}(\mu, \sigma)$. Then the likelihood function can be expressed as:
\begin{equation}
    \mathcal{L} (D|\boldsymbol{\theta}) = \prod^n_{i=1} f(\nu_i | \boldsymbol{\theta}) 
\end{equation}
where the index $i$ runs through all observed points in the spectrum. The explicit expression takes the form: 
\begin{equation}
\mathcal{L} (D|\boldsymbol{\theta}) = \frac{1}{(2\pi\sigma^2)^{n/2}}
    \exp \left(
    -\frac{1}{2} \sum^n_{i=1} \frac{(D(\nu_i) - S_T(\nu_i | \boldsymbol{\theta}))^2}{\sigma^2}
    \right)
\end{equation}
where $\sigma$ corresponds to the uncertainty value of the observations. Ideally, one should use the covariance matrix which is not always available in spectroscopic studies. We estimate the uncertainty $\sigma$ using the standard deviation of the observed data to the fitted continuum.

% =============================================================================
% Physics Performance tests
% =============================================================================
\section{Application on simulated and real spectra}
\label{sec:performance}

In this section we present two scenarios designed to test the accuracy of the model described in Section \ref{sec:physics} and the capability of NIRDust's implementation to recover the true values of its parameters.

We consider the following cases: the use of controlled synthetic spectra, and the real case of NGC 5128 (CenA), which also serves as an independent comparison with the results reported by \citet{Burtscher2015} for this galaxy.

The goal of these tests is to evaluate the impact of the characteristics of the input spectra and the masking procedure on the estimated parameters, particularly the temperature. Moreover, the tests over synthetic spectra intend to serve as a guide to estimate the uncertainties involved when using NIRDust on real spectra.

% SYNTH SPECTRA =========================================================
\subsection{Synthetic spectra: accuracy in the presence of noise}
\label{subsec:synthetic}

For this test we built multiple synthetic models that represent the Target and Reference spectra where we vary the following conditions of interest to asses their impact.

\begin{itemize}
    \item The signal-to-noise (S/N) ratio of both spectra.
    \item The percentage and location of points removed when masking the spectral features.
\end{itemize}

We simultaneously consider a wide range of temperature values, ranging between 500 and 1600\,K. For simplicity we modeled the stellar population of the Reference Spectrum as a decreasing linear function. The Target spectrum was then modeled according to equation \ref{eq2} where we adopted fixed parameter values of $\alpha = 3.5$, $\beta= 8.3$, and $\gamma = -3.3$. However, we have studied the behaviour of the results presented here when these parameters are varied along the default bounds range for fit\_blabckbody()  and found that the results remain unaltered. In all cases a different gaussian noise (different random seed) is added to each spectra to obtain the desired S/N ratio. The spectral resolution of both spectra is set at 3.5 \AA /pix and they consist of 541 spectral points. Throughout all this work, we've consistently measured the S/N ratio using NIRDust's \texttt{compute\_noise} class method. This method calculates the noise by determining the standard deviation of the continuum-subtracted spectrum within a user-defined wavelength interval. This interval must be carefully chosen to represent the noise characteristics of the entire spectrum. Subsequently, the mean signal within this interval is divided by the calculated noise value yielding an average S/N ratio per pixel.

\begin{figure*}

    \includegraphics[width=1\linewidth]{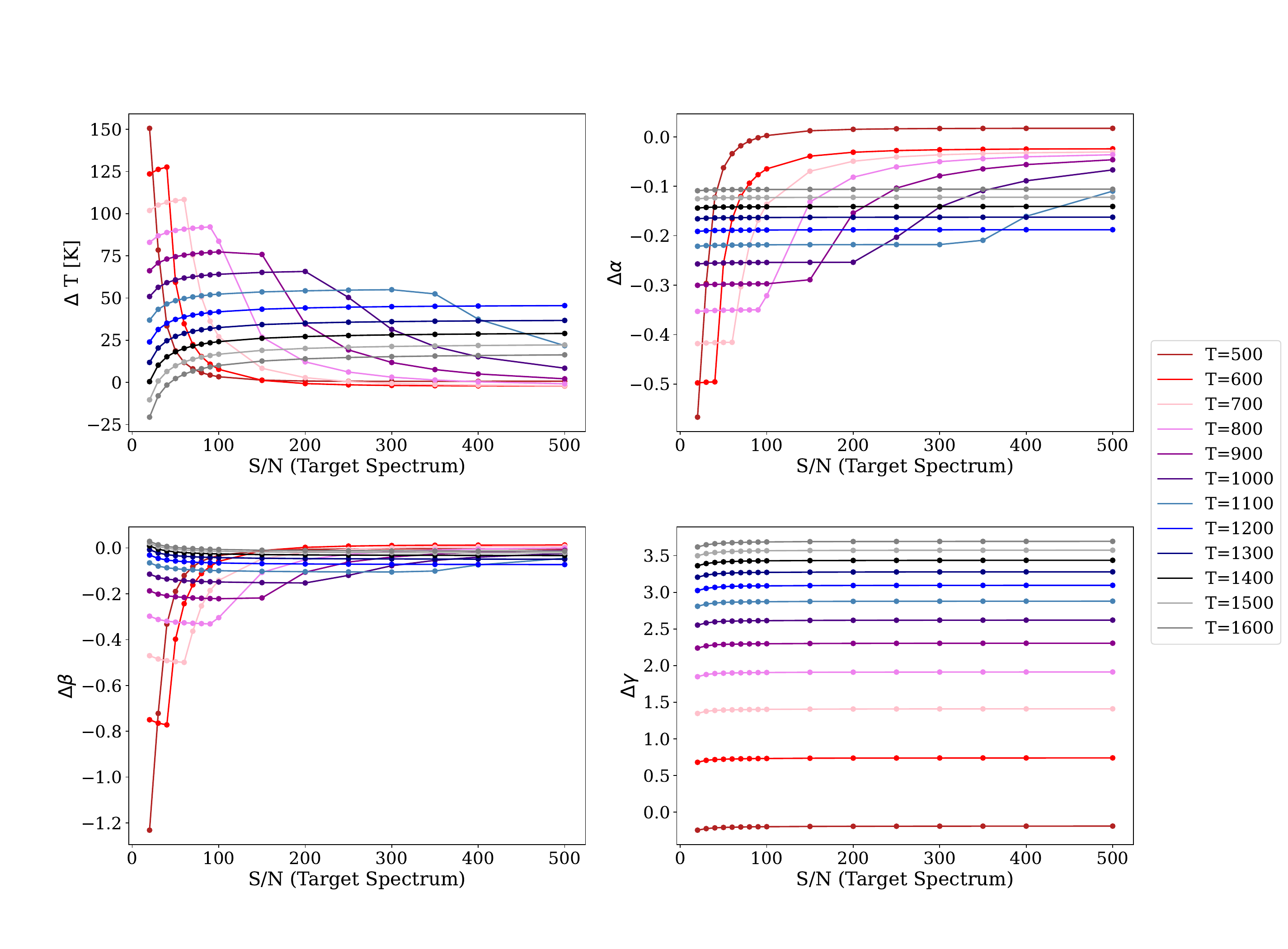}

\caption{Results of the fitting procedure when varying the S/N of both the Target and the Reference spectra for different temperatures. The legend shows the temperatures of the blackbody used to construct the synthetic Target Spectrum in each case. The vertical axis in all panels shows the respective difference of the estimated value of a parameter minus its true value. The horizontal axis corresponds to the S/N ratio of the Target Spectrum that is equal to the S/N ratio of the Reference Spectrum in this case.}
\label{fig:snr}

\end{figure*}

%  FIGURE 1 DESCRIPTION ==================================================
In Fig \ref{fig:snr} the results for the first test are presented. In this case the S/N of both synthetic spectra are equal and varied from 20 to 500 for each value of temperature, with a higher sampling for S/N $< 100$. The fitting procedure was run with higher number of iterations (niter=700) instead of the default value (niter=200) of iterations to achieve convergence in the low S/N points. The deviation of the fitted temperature from the real value, $\Delta T$, for different S/N is shown in the top left panel.  In the top right, bottom left and bottom right panels the variations for the $\alpha$, $\beta$, and $\gamma$ parameters are respectively presented. At high S/N ($\gtrsim 150$) the temperature is accurately estimated for temperatures below 900\,K, but for higher temperatures a systematic and constant overestimation of $\lesssim$ 50\,K is obtained. However, with decreasing S/N (20-50), lower temperatures show an increase in the overestimation reaching $\Delta T \sim$ 100\,K. As can be seen in the other three panels, $\alpha$ and $\beta$ follow the same trend, remaining constant for S/N $\gtrsim$ 150 and deviating for lower S/N in the cases of the lower temperatures ($\lesssim$ 900\,K). The $\gamma$ parameter remains constant along the entire T and S/N ranges and represents a minor contribution to the total flux ($<1\%$).

Summarizing, for all values of temperature we find that the uncertainty remains below 150\,K with a tendency to decrease for higher temperatures, as one would intuitively expect. To achieve an uncertainty below 50\,K a Target Spectrum with S/N$\gtrsim$ 150 is required. It is important to remark that in all cases with S/N $\geq$ 20 the uncertainty correspond to a systematic overestimation. The user can consider this as a "correction factor" that can be applied to the obtained temperature. An example of this correction is presented in Subsec. \ref{subsec:comparison}.

Three additional cases where the S/N of the Reference Spectrum is lower than the S/N of the Target Spectrum are presented in Appendix \ref{appendix}.

\begin{figure*}

    \includegraphics[width=1\linewidth]{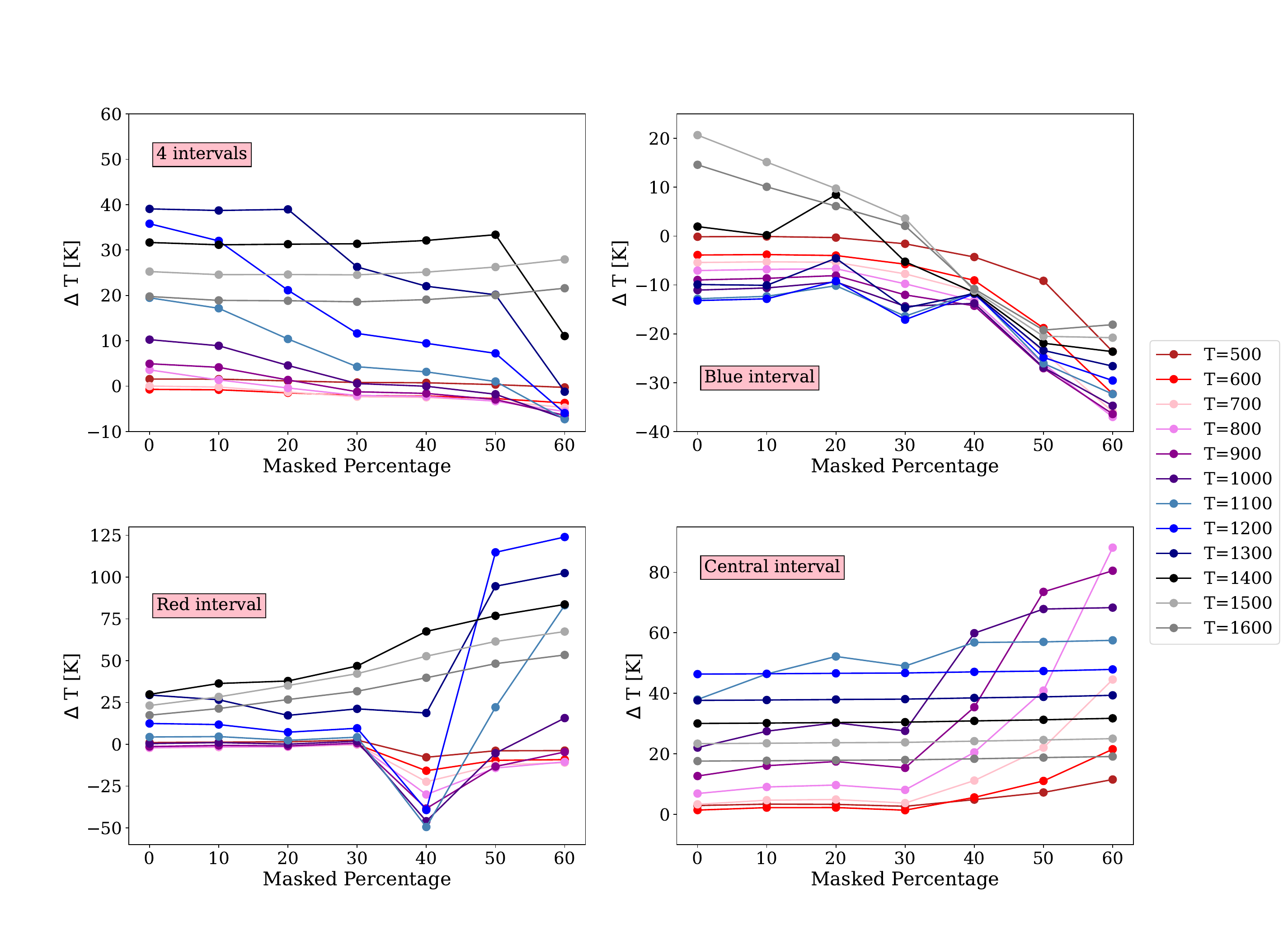}

\caption{Uncertainty in the estimated temperature when varying the percentage of points removed from both the Target and the Reference spectra for different temperatures using \texttt{mask\_spectrum()}. Four masks were implemented: 4 equi-distant and equi-sized intervals distributed along the spectral range (top left panel); one interval in the blue part of the spectrum at 2.07\,$\mu$m (top right panel); one interval in the red part of the spectrum at 2.24\,$\mu$m (bottom left panel); and one interval in the central part of the spectral range at 2.15\,$\mu$m (bottom right panel). In all cases the percentage of removed points is the sum of the masked intervals. The temperatures showed in the legend are the temperatures of the blackbody model of the synthetic Target Spectrum in each case.}
\label{fig:mask}

\end{figure*}

%  FIGURE 2 DESCRIPTION ==================================================
In Fig. \ref{fig:mask} we present results regarding the application of masks used to remove spectral features or high noise regions. The test consisted on computing the difference between the fitted temperature and the real temperature of the synthetic spectrum for different masking scenarios. Through this analysis, our objective was to discern the presence of spectral regions and/or a minimum of available spectral points crucial for accurate dust temperature determination, emphasizing the necessity for meticulous consideration when implementing masking strategies. The synthetic spectra are the same as used in the previous test but the S/N is fixed at a high value of 800 to minimize the uncertainties introduced by the gaussian noise.

Four cases where considered: 
\begin{itemize}
\item "4 intervals" mask: a mask consisting of 4 equi-distant and equi-sized intervals distributed along the spectral range (top left panel).

\item "Blue interval" mask: a mask consisting of one interval in the blue part of the spectrum at 2.07\,$\mu$m (top right panel). 

\item "Red interval" mask: a mask consisting of one interval in the red part of the spectrum at 2.24\,$\mu$m (bottom left panel). 

\item "Central interval" mask: and a mask consisting in one interval in the central part of the spectral range at 2.15\,$\mu$m (bottom right panel). In all cases the percentage of removed points is the sum of the masked intervals and increases between 0 and 60\%. 

\end{itemize}

In all cases the uncertainty in the estimated temperature is mainly constant until a 30\% of mask percentage is reached. Taking into account that the S/N of the spectra involved in this test is of 800 it can be considered that this initial overestimation of the temperature is attributable to the S/N effect as analyzed in the previous test. As the masking percentage increases, however, the uncertainty in the obtained temperature changes in all the cases except the 4 intervals mask (top left panel). In the other three cases, the uncertainty increases and even changes sign. This is expected since the 4 intervals mask is the one that keeps higher amounts of points distributed along the spectral range and hence is expected to retain the information of the continuum slope. Therefore, if the mask involved consists of one large interval placed anywhere along the spectral range it is not recommended to mask beyond 30\% of the points.

% REAL SPECTRA =========================================================
\subsection{Real spectra: Centaurus A}
\label{subsec:comparison}

We have used three different sets of spectra of NGC 5128 (Centaurus A), obtained with different instruments and focal plane sampling techniques, to measure the hot dust temperature and compare it to the temperature reported by \citet{Burtscher2015} 
of $(796 \pm 1)$\,K for the hot dust in Cen\,A nucleus measured in a 1" diameter aperture. Their spectra were selected from the ESO SINFONI \citep{Eisenhauer2003} database and correspond to IFU observations in K-band with seeing of 0.6" taken in March-April, 2005 \citep{Neumayer2007}. 
NGC 5128 was chosen as the primary subject of our comparative study due to the availability of two extra suitable datasets for direct analysis obtained with Flamingos-2 and GNIRS both at the Gemini Observatory. Additionally, Burtscher et al. kindly provided us the SINFONI cube of NGC 5128 they used in their work and the spectrum of the star they used as Reference Spectrum for the stellar population HD 176617 (private communication). We adopted a distance of 3.84 Mpc for NGC 5128 \citep{Rejkuba2004} to calculate the linear radii in pc.

For this comparative analysis we used two different approaches for the Reference Spectrum in order to compare them: in one hand an off-nuclear spectrum of the same galaxy as described in Sec.\ref{subsec:reference spectrum} (Reference 1), and in the other hand, the spectrum of the star used in \cite{Burtscher2015} (Reference 2).

We extracted Target and Reference spectra from our two datasets: 

\begin{itemize}
    \item $K_{long}$ longslit spectra taken with Flamingos-2 \citep{Eikenberry2008, Gomez2012} at Gemini South with PA =151\textdegree \ (semimajor-axis direction), on February 21, 2021. The percentage of masking implemented is of 30\% and the mask consist of 8 narrow intervals distributed along the band with bigger presence in the blue part of the spectra. Taking this into account and considering the behaviours of similar masks presented in Fig. \ref{fig:mask} (the cases of masks composed of 4 intervals and one blue interval) the uncertainty is expected to be dominated by the S/N of the spectra (see subsec. \ref{subsec:synthetic}). 
    \vspace{0.1cm}
    
    \item A $K_{long}$ data cube taken with GNIRS \citep{Elias2006} at Gemini South in 2005 \citep[][ Program ID GS-2005A-Q-38]{Diaz2021}. The percentage of masking is of 10\% and the mask consist of one interval in the red part of the spectrum. For this kind of mask, the uncertainty is expected to be only attributed to the S/N. 
\end{itemize}

\begin{figure}

    \includegraphics[width=0.9\linewidth]{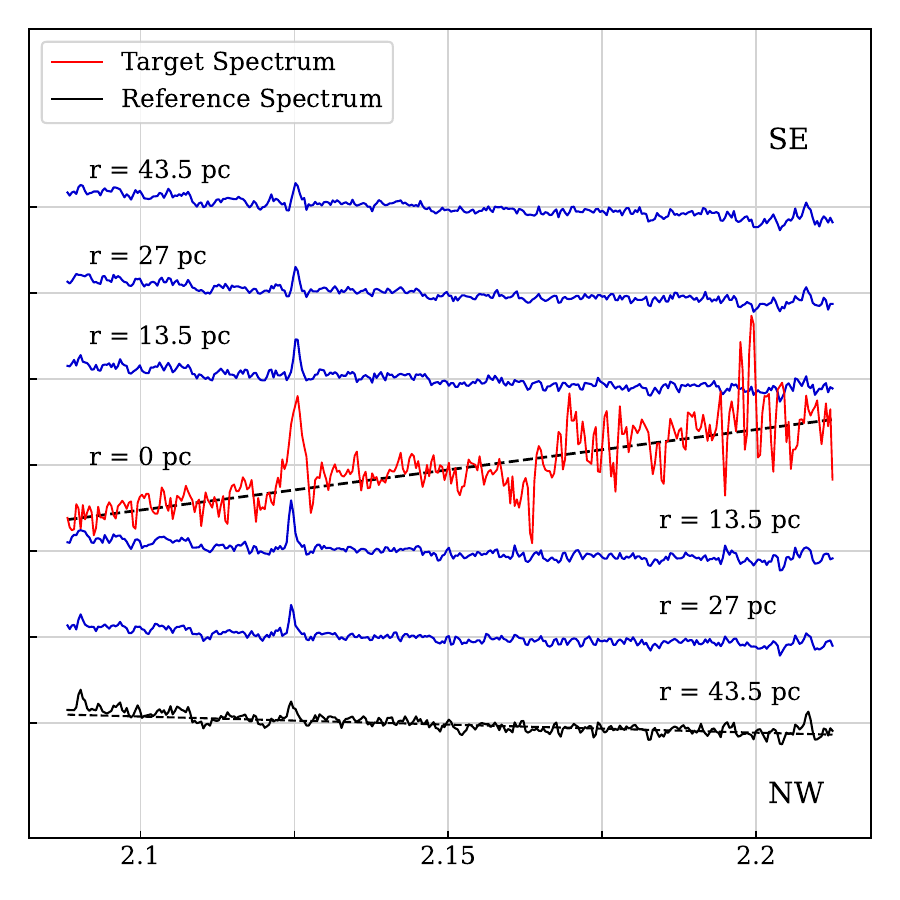}
    \includegraphics[width=0.9\linewidth]{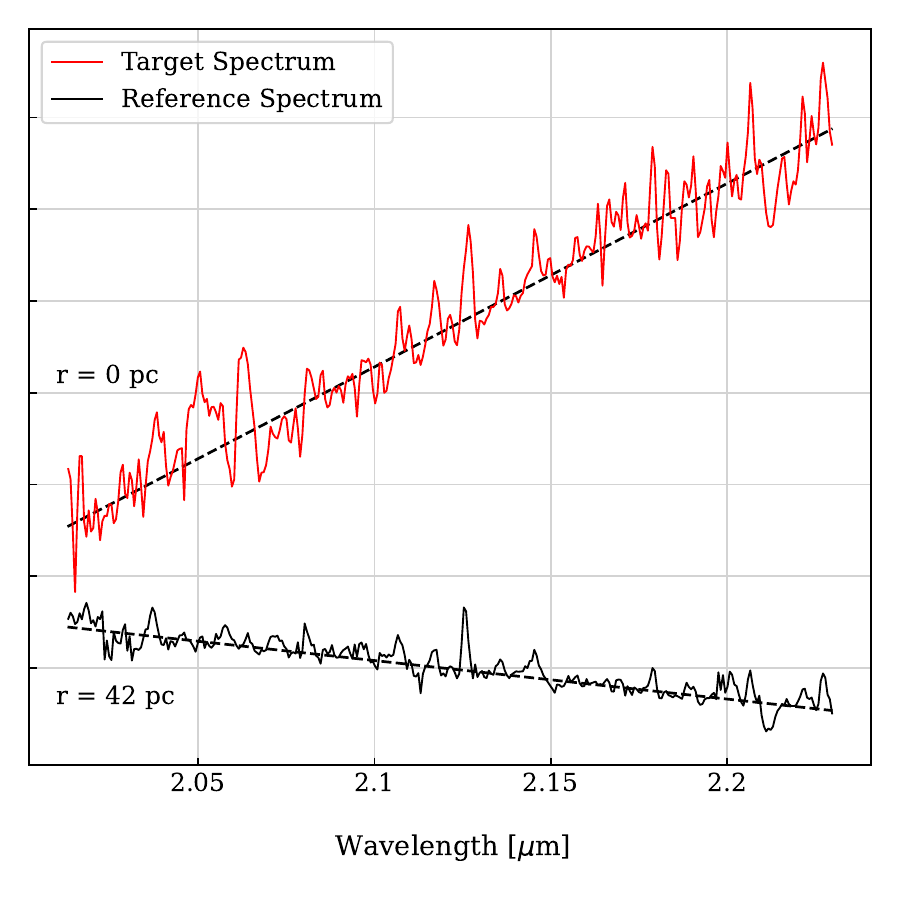}

\caption{Top panel: Flamingos-2 spectra extracted at different radii (r) at both sides of the slit (SE and NW directions). In red and black the Target and Reference spectra are highlighted. Bottom panel: GNIRS spectra extracted at the nucleus (Target) and at 42 pc of radius (Reference), in this case the Reference Spectrum is a combination of spectra extracted at 42 pc in several regions of the IFU field, for more details see the text. The dotted black line is a guide to compare the Target and Reference spectra slopes.
}
\label{fig:spectra}

\end{figure}

In the top panel of Fig. \ref{fig:spectra}, the Flamingos-2 spectra are displayed for different radii, spanning the range from 0 to 43 pc, on both sides of the slit. The nuclear spectrum, designated as the Target Spectrum, is highlighted in red, while the Reference Spectrum is depicted in black. Notably, the blue spectra extracted from various radii exhibit a consistent slope of the spectral continuum across all spectra except the nuclear one. This means that the excess punitively produced by the hot dust emission is present only in the nuclear spectrum and hence all the other blue spectra displayed in the Figure have the potential to serve as a reliable Reference Spectrum. Our choice, however, was directed towards one of the more distant spectra, thereby ensuring a substantial separation from the nucleus. For the GNIRS dataset, shown in the bottom panel of Fig. \ref{fig:spectra}, a distinct approach was necessary due to the instrument's limited field of view (FOV). Instead of employing annular apertures, circular apertures were extracted from different regions surrounding the nucleus at a fixed distance of 2.25". Integrating these circular aperture extractions yielded the Reference Spectrum. As a result of this methodology, Fig. \ref{fig:spectra} exclusively showcases the Target Spectrum and the Reference Spectrum. It's imperative to note that the validation of the Reference Spectrum's integrity was supported by the Flamingos-2 spectra observations, ensuring the absence of red excess beyond the nucleus.

The characteristics of the four extracted spectra, along with the SINFONI spectra from \cite{Burtscher2013}, are summarized in Table \ref{table:data}. The table includes information on the masking percentage, the temperature obtained using NIRDust, and the expected uncertainty calculated from the S/N the two spectra involved in each fit.
The table is divided into three parts corresponding to the instrument used to capture the Target Spectrum. For the Flamingos-2 and GNIRS data, there are two Reference Spectra: one extracted from the same galaxy (reference 1) and the SINFONI spectrum of HD 176617 (reference 2). A good fit for the SINFONI Target Spectrum could not be obtained using a Reference Spectrum extracted from the same data cube. Therefore, this result is not included in the table.

\begin{table*}
\begin{center}
\caption{Description of the three data sets used to measure the dust temperature with NIRDust. The obtained results are presented in column 7. The masking percentage is only indicated for the Reference Spectrum as may not be the same for all the references. The Target Spectrum presents the same masking percentage as the Reference in each case.}
\begin{tabular}{lccccccc}
\toprule 
\toprule 
 Spectrum  & data type & S/N & aperture  & radius of  & masking & temperature & expected \\
 & &  & diameter ["] & extraction ["] &  percentaje & [K] &$\Delta$T \\
\toprule 
 & & & Flamingos-2 & &  & \\
\midrule
Target & longslit & 95 & 1 & 0 & - & - & - \\
Reference 1 & longslit & 72 & 1.08 & 3.34 & 30 & 662 & 25 - 85 \\
Reference 2 (HD 176617) & IFU & 40 & point source & - & 30 & 667 & 90 - 110\\
\midrule
 & & & GNIRS & & & \\
\midrule
Target & IFU  & 87 & 1.05 & - & - & - & - \\ 
Reference 1 & IFU & 74 & 1.05 & 2.25 & 10 & 697 & 5 - 75 \\
Reference 2 (HD 176617) & IFU & 40 & point source & - & 0 & 607 & 90-110 \\
\midrule
 & & & SINFONI & & & \\
\midrule
Target & IFU  & 205 & 1. & 0 & - & - & -\\ 
Reference (HD 176617) & IFU & 40 & point source & - & 30 & 649 & 100 - 110 \\
\bottomrule 
\label{table:data}
\end{tabular}
\end{center}
\end{table*}

\begin{figure*}

    \includegraphics[width=0.48\linewidth]{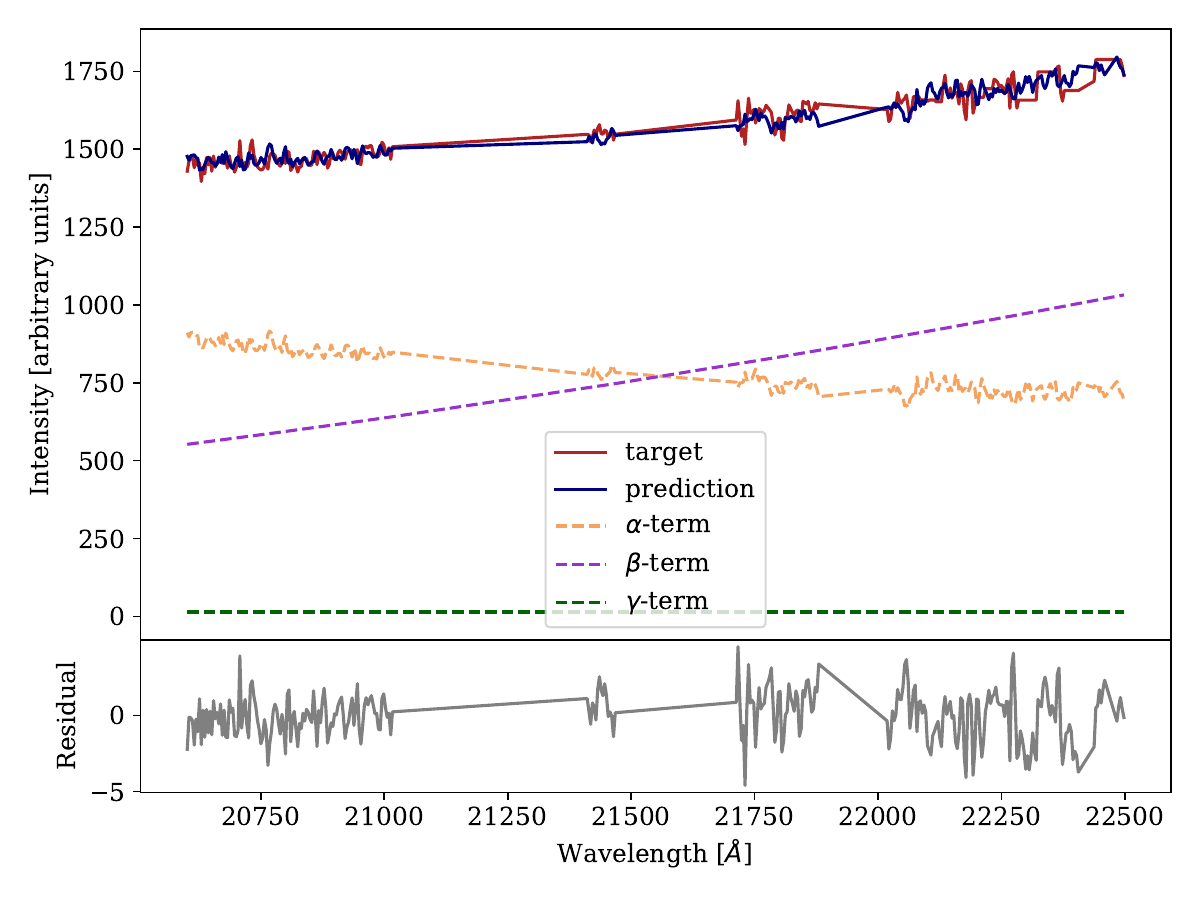}
    \includegraphics[width=0.48\linewidth]{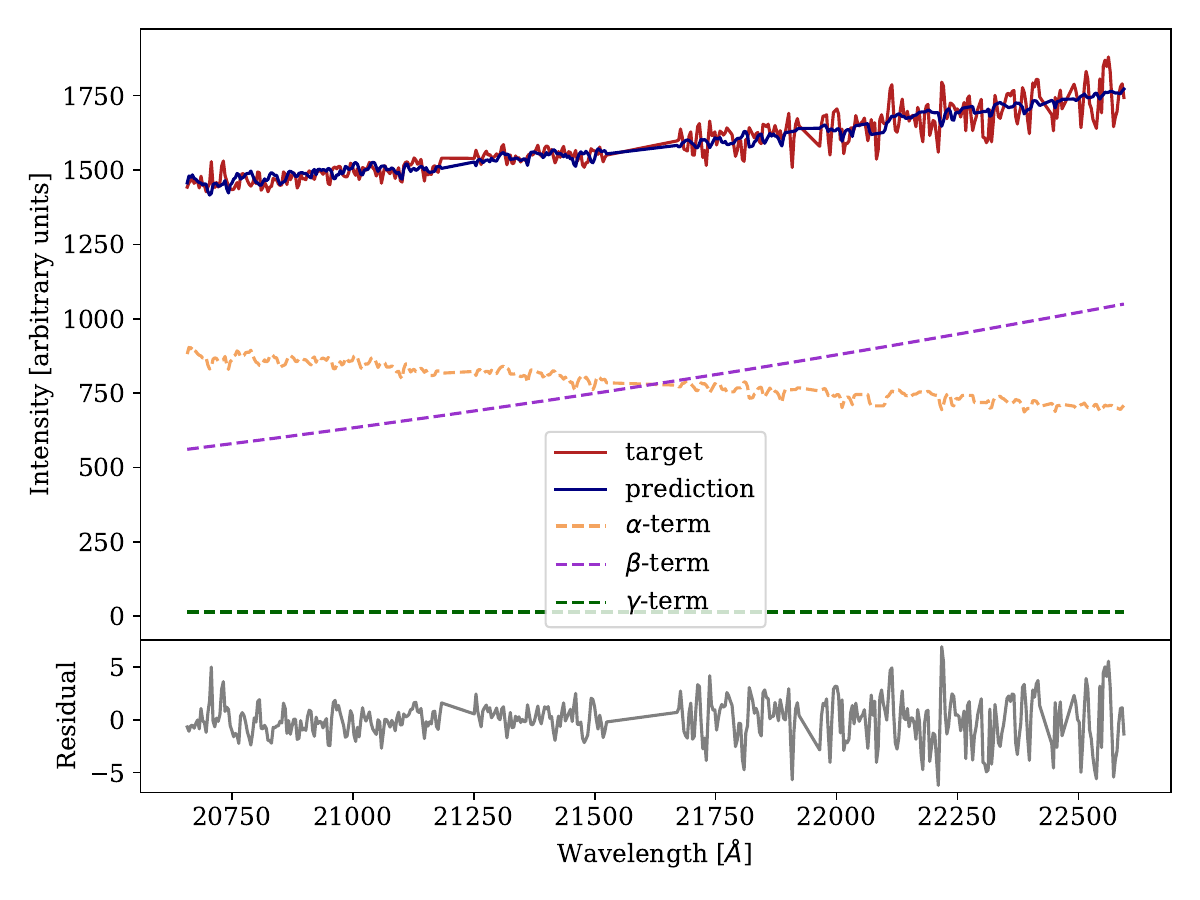}
    \includegraphics[width=0.48\linewidth]{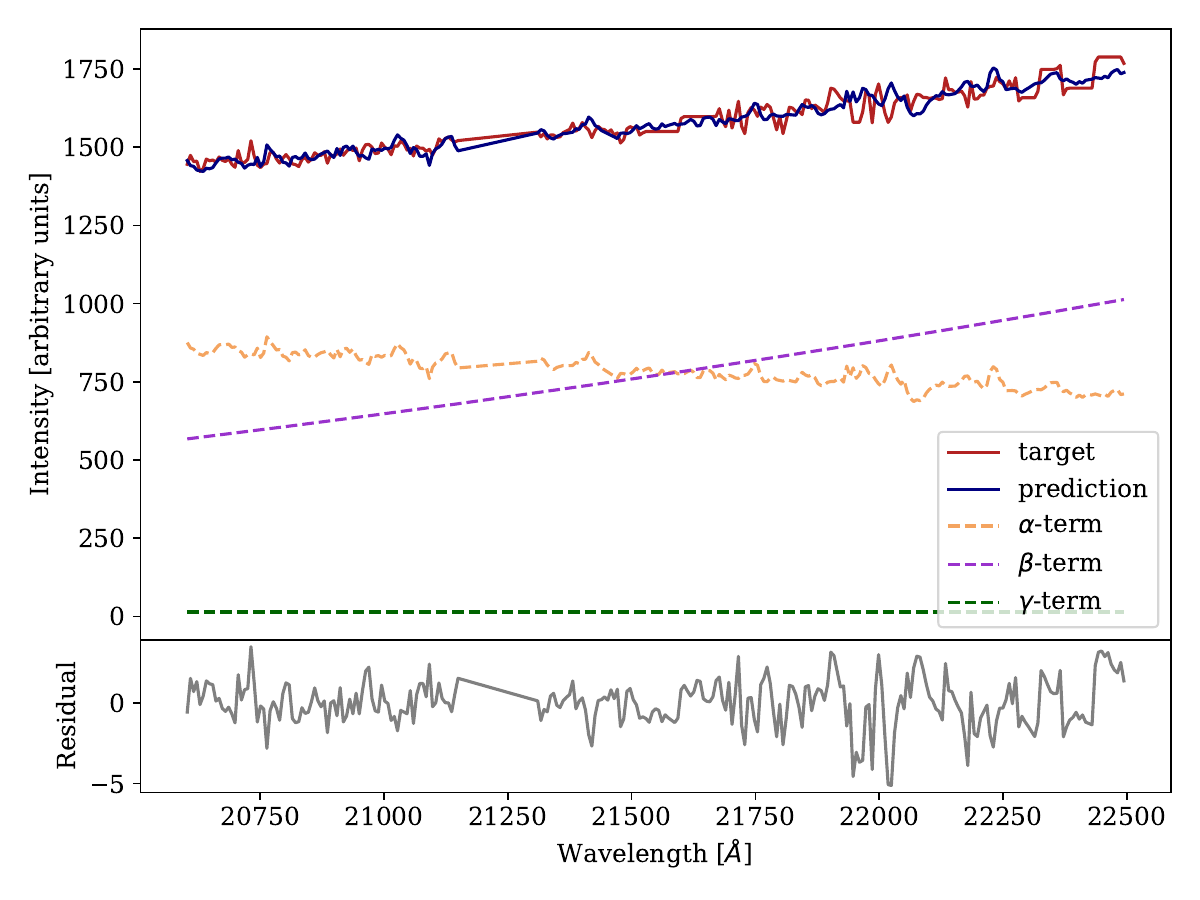}
    \includegraphics[width=0.48\linewidth]{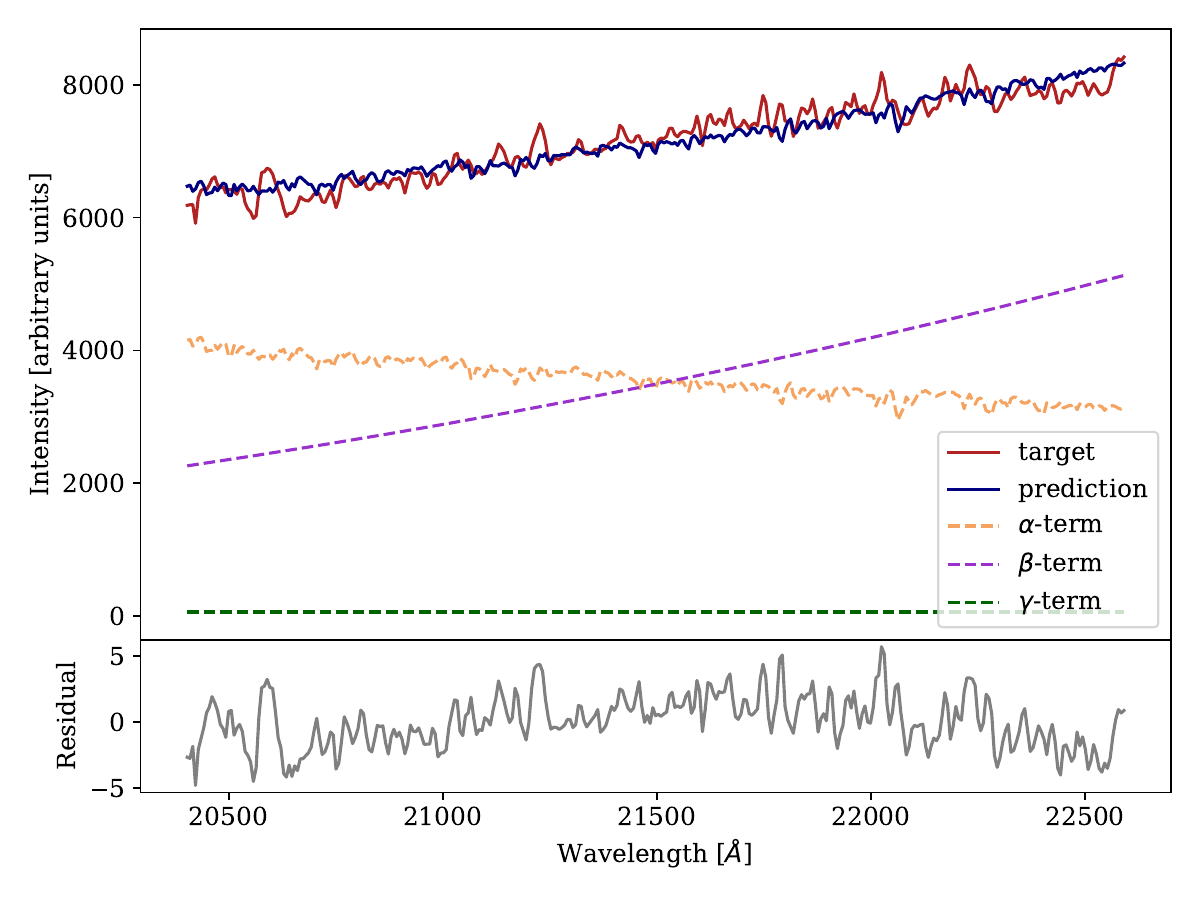}
    \includegraphics[width=0.48\linewidth]{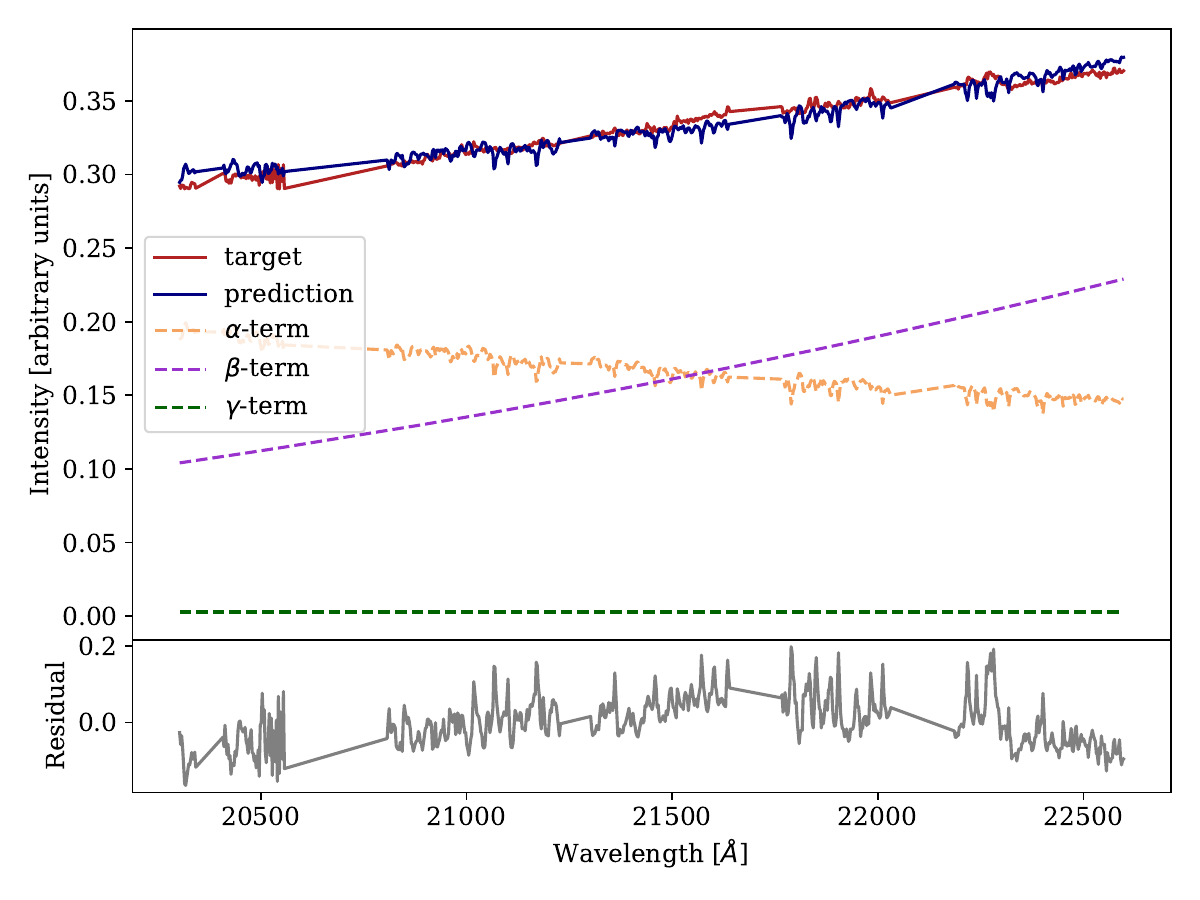}

\caption{NIRDust fitting results for the 5 combinations of spectra described in Table \ref{table:data}. First row corresponds to the Flamingos-2 Target Spectra, the second row to the GNIRS Target Spectra and the third row to the SINFONI Target Spectra. For the first two rows the left column corresponds to the Reference Spectrum extracted from the same galaxy and the right column corresponds to the cases where the spectrum of HD 176617 was used as reference. In the 5 cases the plots are the actual output of NIRDust, to see the obtained temperatures reference to Table \ref{table:data}}
\label{fig:results}

\end{figure*}

In Fig. \ref{fig:results} we present the NIRDust fitting results for the 5 combinations of spectra described in Table \ref{table:data}.
We obtained temperatures of 662\,K (Reference 1) and 667\,K (Reference 2) from the Flamingos-2 spectrum and of 697\,K (Reference 1) and 607\,K (Reference 2) from the GNIRS data cube. As previously mentioned, the synthetic spectra tests of Sec.\ref{subsec:synthetic} showed that the S/N of the spectra causes an overestimation of the temperature. We present the overestimation of the temperature estimated for each pair of spectra in column 8 of Table \ref{table:data}. Sec.\ref{subsec:synthetic}. For the SINFONI spectrum, the obtained dust temperature is 649\,K, which is 147\,K lower than the temperature of 796\,K reported by \cite{Burtscher2015}. Nevertheless, all temperatures obtained with NIRDust are consistent with each other within a typical uncertainty of $\sim$ 100\,K.
Since both the Target and Reference spectra from SINFONI are the same as in \cite{Burtscher2015}, this difference in temperature may arise from differences in the methodology, specifically the equation for the linear combination used to perform the fitting. Despite this difference is worth noting that all dust temperatures measured by NIRDust and by \cite{Burtscher2015} fall in the range of relatively low temperature that have been previously reported for Seyfert 2 nucleii such as the one hosted NGC 5128 \citep{Riffel2009}.

More extensive experiments on the impact of the Reference Spectrum using different options and data sets are necessary in order to further constrain the uncertainties involved in the estimation of the dust temperature using the method described in Sec. \ref{sec:physics}.

% =============================================================================
% Conclusions
% =============================================================================
\section{Summary and future work}
\label{sec:conclusions}

In this work we have presented an analysis on the proposed model to describe the NIR emission of hot dust in Type 2 AGNs. We designed a set of multiple synthetic spectra that allowed a thorough analysis on the behaviour of the mathematical model that describes the physics of the emitting dust. This study provided useful insights at the moment of analysing real spectra from Flamingos-2 (Gemini South), GNIRS (Gemini North) and SINFONI (VLT).

The results presented here can be summarized in the following key points:
\begin{itemize}
    \item The parametrization choice of the model in terms of $\boldsymbol{\theta} = \{T, \alpha, \beta, \gamma\}$ and a basin-hopping optimization algorithm are suitable to detect the hot dust component in real spectra.
    \item The tests performed over synthetic spectra show that  estimated temperature presents a systematic error which value depends primarily on the S/N of the Target Spectrum and can be accurately characterized for each dataset. In most cases the error  of the temperatures is below 100\,K for S/N $\geq$ 20; for the extreme cases of low T and low S/N the error can rise up to 150\,K.
    \item The five dust temperatures obtained for NGC 5128 (Centaurus A) using three independent data sets from Flamingos-2 (662\,K and 667\,K), GNIRS (697\,K and 607\,K) and SINFONI (653\,K) and two different Reference Spectra for the first two data sets are consistent within a typical uncertainty of $\sim$ 100\,K. The obtained temperatures for the three data sets fall in the low temperature regime for Type 2 AGN NIR emitting dust.
    
    \item The comparison of the dust temperatures obtained with NIRDust for the SINFONI spectrum used by \citet{Burtscher2015} for NGC 5128 yields a difference of 147\,K with their reported temperature of 796\,K.
\end{itemize}

To achieve these results we developed NIRDust, an open source Python package that uses K-band rest frame spectra to detect hot dust in Type 2 AGNs and measure its temperature. This package is open to the astronomical community and is an effort to provide a standardized and reproducible procedure to measure the temperature of hot dust present in these objects. The performance of NIRDust in the presence of gaussian noise is satisfactory and the uncertainties involved can be estimated.

As a continuation of this work we intend to characterize the impact of different Reference Spectrum choices on the estimated temperature. New features will be added, like an optional power law function in the model to account for any scattered disk accretion emission present in the Target Spectrum.

\section*{Acknowledgements}

The authors would like to thank Leonard Burtscher and Orban De Xivry Gilles for generously providing the NGC 5128 data they used in their work as it significantly contributed to our analysis; to their families and friends, OAC and IATE astronomers for useful comments and suggestions and the anonymous referee for a deep reading and useful suggestions.
This work was partially supported by the Consejo Nacional
de Investigaciones Cient\'ificas y T\'ecnicas (CONICET, Argentina) and the Secretar\'ia de Ciencia y Tecnolog\'ia de la Universidad Nacional de C\'ordoba (SeCyT-UNC, Argentina), and grant PICT 2017-3301 awarded by Foncyt.
G.G, J.A., J.B.C and M.Ch. are supported by a fellowship from CONICET. D.M. is a member of the Carrera de Investigador Cient\'\i fico of CONICET.
This research has made use of the
\url{http://adsabs.harvard.edu/}, Cornell University xxx.arxiv.org repository, adstex (\url{https://github.com/yymao/adstex}), astropy and the Python programming language.

%%%%%%%%%%%%%%%%%%%%%%%%%%%%%%%%%%%%%%%%%%%%%%%%%%
\section*{Data Availability}

The code used within this article is fully available at \url{https://github.com/Gaiana/nirdust} along with some example data. The raw GNIRS data are available from the Gemini Observatory Archive (https://archive.gemini.edu) with program ID GS-2005A-Q-38.


\bibliographystyle{mnras}
\bibliography{main} % if your bibtex file is called example.bib

\begin{thebibliography}{}
\makeatletter
\relax
\def\mn@urlcharsother{\let\do\@makeother \do\$\do\&\do\#\do\^\do\_\do\%\do\~}
\def\mn@doi{\begingroup\mn@urlcharsother \@ifnextchar [ {\mn@doi@} {\mn@doi@[]}}
\def\mn@doi@[#1]#2{\def\@tempa{#1}\ifx\@tempa\@empty \href {http://dx.doi.org/#2} {doi:#2}\else \href {http://dx.doi.org/#2} {#1}\fi \endgroup}
\def\mn@eprint#1#2{\mn@eprint@#1:#2::\@nil}
\def\mn@eprint@arXiv#1{\href {http://arxiv.org/abs/#1} {{\tt arXiv:#1}}}
\def\mn@eprint@dblp#1{\href {http://dblp.uni-trier.de/rec/bibtex/#1.xml} {dblp:#1}}
\def\mn@eprint@#1:#2:#3:#4\@nil{\def\@tempa {#1}\def\@tempb {#2}\def\@tempc {#3}\ifx \@tempc \@empty \let \@tempc \@tempb \let \@tempb \@tempa \fi \ifx \@tempb \@empty \def\@tempb {arXiv}\fi \@ifundefined {mn@eprint@\@tempb}{\@tempb:\@tempc}{\expandafter \expandafter \csname mn@eprint@\@tempb\endcsname \expandafter{\@tempc}}}

\bibitem[\protect\citeauthoryear{{Allen} \& {Schmidt}}{{Allen} \& {Schmidt}}{2015}]{Allen2015}
{Allen} A.,  {Schmidt} J.,  2015, \mn@doi [Journal of Open Research Software] {10.5334/jors.bv}, \href {https://ui.adsabs.harvard.edu/abs/2015JORS....3E..15A} {3, E15}

\bibitem[\protect\citeauthoryear{{Alonso-Herrero}, {Simpson}, {Ward}  \& {Wilson}}{{Alonso-Herrero} et~al.}{1998}]{AlonosoHerrero1998}
{Alonso-Herrero} A.,  {Simpson} C.,  {Ward} M.~J.,   {Wilson} A.~S.,  1998, \mn@doi [\apj] {10.1086/305269}, \href {https://ui.adsabs.harvard.edu/abs/1998ApJ...495..196A} {495, 196}

\bibitem[\protect\citeauthoryear{{Alonso-Herrero} et~al.,}{{Alonso-Herrero} et~al.}{2018}]{Alonso-Herrero2018}
{Alonso-Herrero} A.,  et~al., 2018, \mn@doi [\apj] {10.3847/1538-4357/aabe30}, \href {https://ui.adsabs.harvard.edu/abs/2018ApJ...859..144A} {859, 144}

\bibitem[\protect\citeauthoryear{{Antonucci}}{{Antonucci}}{1993}]{Antonucci1993}
{Antonucci} R.,  1993, \mn@doi [\araa] {10.1146/annurev.aa.31.090193.002353}, \href {http://adsabs.harvard.edu/abs/1993ARA\%26A..31..473A} {31, 473}

\bibitem[\protect\citeauthoryear{{Asmus}}{{Asmus}}{2019}]{Asmus2019}
{Asmus} D.,  2019, \mn@doi [\mnras] {10.1093/mnras/stz2289}, \href {https://ui.adsabs.harvard.edu/abs/2019MNRAS.489.2177A} {489, 2177}

\bibitem[\protect\citeauthoryear{{Asmus}, {H{\"o}nig}  \& {Gandhi}}{{Asmus} et~al.}{2016}]{Asmus2016}
{Asmus} D.,  {H{\"o}nig} S.~F.,   {Gandhi} P.,  2016, \mn@doi [\apj] {10.3847/0004-637X/822/2/109}, \href {https://ui.adsabs.harvard.edu/abs/2016ApJ...822..109A} {822, 109}

\bibitem[\protect\citeauthoryear{{Astropy Collaboration} et~al.,}{{Astropy Collaboration} et~al.}{2013}]{Astropy}
{Astropy Collaboration} et~al., 2013, \mn@doi [\aap] {10.1051/0004-6361/201322068}, \href {https://ui.adsabs.harvard.edu/abs/2013A&A...558A..33A} {558, A33}

\bibitem[\protect\citeauthoryear{{Burtscher} et~al.,}{{Burtscher} et~al.}{2013}]{Burtscher2013}
{Burtscher} L.,  et~al., 2013, \mn@doi [\aap] {10.1051/0004-6361/201321890}, \href {https://ui.adsabs.harvard.edu/abs/2013A&A...558A.149B} {558, A149}

\bibitem[\protect\citeauthoryear{{Burtscher} et~al.,}{{Burtscher} et~al.}{2015}]{Burtscher2015}
{Burtscher} L.,  et~al., 2015, \mn@doi [\aap] {10.1051/0004-6361/201525817}, \href {https://ui.adsabs.harvard.edu/abs/2015A&A...578A..47B} {578, A47}

\bibitem[\protect\citeauthoryear{{Combes} et~al.,}{{Combes} et~al.}{2019}]{combes2019}
{Combes} F.,  et~al., 2019, \mn@doi [\aap] {10.1051/0004-6361/201834560}, \href {https://ui.adsabs.harvard.edu/abs/2019A&A...623A..79C} {623, A79}

\bibitem[\protect\citeauthoryear{{D{\'\i}az} et~al.,}{{D{\'\i}az} et~al.}{2021}]{Diaz2021}
{D{\'\i}az} R.~J.,  et~al., 2021, Boletin de la Asociacion Argentina de Astronomia La Plata Argentina, \href {https://ui.adsabs.harvard.edu/abs/2021BAAA...62..219D} {62, 219}

\bibitem[\protect\citeauthoryear{{Dumont}, {Seth}, {Strader}, {Greene}, {Burtscher}  \& {Neumayer}}{{Dumont} et~al.}{2020}]{Dumont2020}
{Dumont} A.,  {Seth} A.~C.,  {Strader} J.,  {Greene} J.~E.,  {Burtscher} L.,   {Neumayer} N.,  2020, \mn@doi [\apj] {10.3847/1538-4357/ab5798}, \href {https://ui.adsabs.harvard.edu/abs/2020ApJ...888...19D} {888, 19}

\bibitem[\protect\citeauthoryear{{Durr{\'e}} \& {Mould}}{{Durr{\'e}} \& {Mould}}{2018}]{Durre2018}
{Durr{\'e}} M.,  {Mould} J.,  2018, \mn@doi [\apj] {10.3847/1538-4357/aae68e}, \href {https://ui.adsabs.harvard.edu/abs/2018ApJ...867..149D} {867, 149}

\bibitem[\protect\citeauthoryear{{Eikenberry} et~al.,}{{Eikenberry} et~al.}{2008}]{Eikenberry2008}
{Eikenberry} S.,  et~al., 2008, in {McLean} I.~S.,  {Casali} M.~M.,  eds,  Society of Photo-Optical Instrumentation Engineers (SPIE) Conference Series Vol. 7014, Ground-based and Airborne Instrumentation for Astronomy II. p. 70140V, \mn@doi{10.1117/12.788326}

\bibitem[\protect\citeauthoryear{{Eisenhauer} et~al.,}{{Eisenhauer} et~al.}{2003}]{Eisenhauer2003}
{Eisenhauer} F.,  et~al., 2003, in {Iye} M.,  {Moorwood} A. F.~M.,  eds,  Society of Photo-Optical Instrumentation Engineers (SPIE) Conference Series Vol. 4841, Instrument Design and Performance for Optical/Infrared Ground-based Telescopes. pp 1548--1561 (\mn@eprint {arXiv} {astro-ph/0306191}), \mn@doi{10.1117/12.459468}

\bibitem[\protect\citeauthoryear{{Elias}, {Joyce}, {Liang}, {Muller}, {Hileman}  \& {George}}{{Elias} et~al.}{2006}]{Elias2006}
{Elias} J.~H.,  {Joyce} R.~R.,  {Liang} M.,  {Muller} G.~P.,  {Hileman} E.~A.,   {George} J.~R.,  2006, in {McLean} I.~S.,  {Iye} M.,  eds,  SPIE Conference Series Vol. 6269, Ground-based and Airborne Instrumentation for Astronomy. p. 62694C, \mn@doi{10.1117/12.671817}

\bibitem[\protect\citeauthoryear{Feigenbaum}{Feigenbaum}{1983}]{feigenbaum1983total}
Feigenbaum A.,  1983, Total quality control

\bibitem[\protect\citeauthoryear{{Ferruit}, {Mundell}, {Nagar}, {Emsellem}, {P{\'e}contal}, {Wilson}  \& {Schinnerer}}{{Ferruit} et~al.}{2004}]{Ferruit2004}
{Ferruit} P.,  {Mundell} C.~G.,  {Nagar} N.~M.,  {Emsellem} E.,  {P{\'e}contal} E.,  {Wilson} A.~S.,   {Schinnerer} E.,  2004, \mn@doi [\mnras] {10.1111/j.1365-2966.2004.08009.x}, \href {http://adsabs.harvard.edu/abs/2004MNRAS.352.1180F} {352, 1180}

\bibitem[\protect\citeauthoryear{Fowler \& Foemmel}{Fowler \& Foemmel}{2006}]{fowler2006continuous}
Fowler M.,  Foemmel M.,  2006, Continuous integration

\bibitem[\protect\citeauthoryear{{G{\'a}mez Rosas} et~al.,}{{G{\'a}mez Rosas} et~al.}{2022}]{Gamez-Rosas2022}
{G{\'a}mez Rosas} V.,  et~al., 2022, \mn@doi [\nat] {10.1038/s41586-021-04311-7}, \href {https://ui.adsabs.harvard.edu/abs/2022Natur.602..403G} {602, 403}

\bibitem[\protect\citeauthoryear{Gamma, Helm, Johnson, Vlissides  \& Patterns}{Gamma et~al.}{1995}]{gamma1995elements}
Gamma E.,  Helm R.,  Johnson R.,  Vlissides J.,   Patterns D.,  1995, Elements of reusable object-oriented software.
 Addison-Wesley Professional Computing Series Vol. 99, Addison-Wesley Reading, Massachusetts

\bibitem[\protect\citeauthoryear{{Garc{\'\i}a-Burillo} et~al.,}{{Garc{\'\i}a-Burillo} et~al.}{2021}]{Garcia-burillo2021}
{Garc{\'\i}a-Burillo} S.,  et~al., 2021, \mn@doi [\aap] {10.1051/0004-6361/202141075}, \href {https://ui.adsabs.harvard.edu/abs/2021A&A...652A..98G} {652, A98}

\bibitem[\protect\citeauthoryear{{Gaspar}, {D{\'\i}az}, {Mast}, {D'Ambra}, {Ag{\"u}ero}  \& {G{\"u}nthardt}}{{Gaspar} et~al.}{2019}]{Gaspar2019}
{Gaspar} G.,  {D{\'\i}az} R.~J.,  {Mast} D.,  {D'Ambra} A.,  {Ag{\"u}ero} M.~P.,   {G{\"u}nthardt} G.,  2019, \mn@doi [\aj] {10.3847/1538-3881/aaf4b9}, \href {https://ui.adsabs.harvard.edu/abs/2019AJ....157...44G} {157, 44}

\bibitem[\protect\citeauthoryear{{Gaspar}, {D{\'\i}az}, {Mast}, {Ag{\"u}ero}, {Schirmer}, {G{\"u}nthardt}  \& {Schmidt}}{{Gaspar} et~al.}{2022}]{Gaspar2022}
{Gaspar} G.,  {D{\'\i}az} R.~J.,  {Mast} D.,  {Ag{\"u}ero} M.~P.,  {Schirmer} M.,  {G{\"u}nthardt} G.,   {Schmidt} E.~O.,  2022, \mn@doi [\aj] {10.3847/1538-3881/ac5ea4}, \href {https://ui.adsabs.harvard.edu/abs/2022AJ....163..230G} {163, 230}

\bibitem[\protect\citeauthoryear{{Glass} \& {Moorwood}}{{Glass} \& {Moorwood}}{1985}]{Glass1985}
{Glass} I.~S.,  {Moorwood} A.~F.~M.,  1985, \mn@doi [\mnras] {10.1093/mnras/214.4.429}, \href {https://ui.adsabs.harvard.edu/abs/1985MNRAS.214..429G} {214, 429}

\bibitem[\protect\citeauthoryear{{Gomez} et~al.,}{{Gomez} et~al.}{2012}]{Gomez2012}
{Gomez} P.~L.,  et~al., 2012, in American Astronomical Society Meeting Abstracts \#219. p. 413.07

\bibitem[\protect\citeauthoryear{{Gratadour}, {Cl{\'e}net}, {Rouan}, {Lai}  \& {Forveille}}{{Gratadour} et~al.}{2003}]{Gratadour2003}
{Gratadour} D.,  {Cl{\'e}net} Y.,  {Rouan} D.,  {Lai} O.,   {Forveille} T.,  2003, \mn@doi [\aap] {10.1051/0004-6361:20031376}, \href {https://ui.adsabs.harvard.edu/abs/2003A&A...411..335G} {411, 335}

\bibitem[\protect\citeauthoryear{{Gravity Collaboration} et~al.,}{{Gravity Collaboration} et~al.}{2020}]{Gravity2020}
{Gravity Collaboration} et~al., 2020, \mn@doi [\aap] {10.1051/0004-6361/201936255}, \href {https://ui.adsabs.harvard.edu/abs/2020A&A...634A...1G} {634, A1}

\bibitem[\protect\citeauthoryear{Holscher, Leifer  \& Grace}{Holscher et~al.}{2010}]{holscher2010read}
Holscher E.,  Leifer C.,   Grace B.,  2010, URL< https://docs. readthedocs. org

\bibitem[\protect\citeauthoryear{{H{\"o}nig}}{{H{\"o}nig}}{2019}]{Honig2019}
{H{\"o}nig} S.~F.,  2019, \mn@doi [\apj] {10.3847/1538-4357/ab4591}, \href {https://ui.adsabs.harvard.edu/abs/2019ApJ...884..171H} {884, 171}

\bibitem[\protect\citeauthoryear{{H{\"o}nig} \& {Kishimoto}}{{H{\"o}nig} \& {Kishimoto}}{2017}]{Honig2017}
{H{\"o}nig} S.~F.,  {Kishimoto} M.,  2017, \mn@doi [\apjl] {10.3847/2041-8213/aa6838}, \href {https://ui.adsabs.harvard.edu/abs/2017ApJ...838L..20H} {838, L20}

\bibitem[\protect\citeauthoryear{{H{\"o}nig}, {Kishimoto}, {Antonucci}, {Marconi}, {Prieto}, {Tristram}  \& {Weigelt}}{{H{\"o}nig} et~al.}{2012}]{Honig2012}
{H{\"o}nig} S.~F.,  {Kishimoto} M.,  {Antonucci} R.,  {Marconi} A.,  {Prieto} M.~A.,  {Tristram} K.,   {Weigelt} G.,  2012, \mn@doi [\apj] {10.1088/0004-637X/755/2/149}, \href {https://ui.adsabs.harvard.edu/abs/2012ApJ...755..149H} {755, 149}

\bibitem[\protect\citeauthoryear{{H{\"o}nig} et~al.,}{{H{\"o}nig} et~al.}{2013}]{Honig2013}
{H{\"o}nig} S.~F.,  et~al., 2013, \mn@doi [\apj] {10.1088/0004-637X/771/2/87}, \href {https://ui.adsabs.harvard.edu/abs/2013ApJ...771...87H} {771, 87}

\bibitem[\protect\citeauthoryear{{Izumi}, {Wada}, {Fukushige}, {Hamamura}  \& {Kohno}}{{Izumi} et~al.}{2018}]{Izumi2018}
{Izumi} T.,  {Wada} K.,  {Fukushige} R.,  {Hamamura} S.,   {Kohno} K.,  2018, \mn@doi [\apj] {10.3847/1538-4357/aae20b}, \href {https://ui.adsabs.harvard.edu/abs/2018ApJ...867...48I} {867, 48}

\bibitem[\protect\citeauthoryear{{Jazayeri}}{{Jazayeri}}{2007}]{Jazayeri2007}
{Jazayeri} M.,  2007, in Future of Software Engineering (FOSE '07). pp 199--213, \mn@doi{10.1109/FOSE.2007.26}

\bibitem[\protect\citeauthoryear{{Kishimoto}, {H{\"o}nig}, {Antonucci}, {Barvainis}, {Kotani}, {Tristram}, {Weigelt}  \& {Levin}}{{Kishimoto} et~al.}{2011}]{Kishimoto2011}
{Kishimoto} M.,  {H{\"o}nig} S.~F.,  {Antonucci} R.,  {Barvainis} R.,  {Kotani} T.,  {Tristram} K.~R.~W.,  {Weigelt} G.,   {Levin} K.,  2011, \mn@doi [\aap] {10.1051/0004-6361/201016054}, \href {https://ui.adsabs.harvard.edu/abs/2011A&A...527A.121K} {527, A121}

\bibitem[\protect\citeauthoryear{{Krolik} \& {Begelman}}{{Krolik} \& {Begelman}}{1988}]{krolik1988}
{Krolik} J.~H.,  {Begelman} M.~C.,  1988, \mn@doi [\apj] {10.1086/166414}, \href {https://ui.adsabs.harvard.edu/abs/1988ApJ...329..702K} {329, 702}

\bibitem[\protect\citeauthoryear{{Leftley}, {Tristram}, {H{\"o}nig}, {Asmus}, {Kishimoto}  \& {Gandhi}}{{Leftley} et~al.}{2021}]{Leftley2021}
{Leftley} J.~H.,  {Tristram} K. R.~W.,  {H{\"o}nig} S.~F.,  {Asmus} D.,  {Kishimoto} M.,   {Gandhi} P.,  2021, \mn@doi [\apj] {10.3847/1538-4357/abee80}, \href {https://ui.adsabs.harvard.edu/abs/2021ApJ...912...96L} {912, 96}

\bibitem[\protect\citeauthoryear{{Lin}, {Zou}, {Kong}, {Lin}, {Mao}, {Cheng}, {Jiang}  \& {Zhou}}{{Lin} et~al.}{2013}]{Lin2013}
{Lin} L.,  {Zou} H.,  {Kong} X.,  {Lin} X.,  {Mao} Y.,  {Cheng} F.,  {Jiang} Z.,   {Zhou} X.,  2013, \mn@doi [\apj] {10.1088/0004-637X/769/2/127}, \href {https://ui.adsabs.harvard.edu/abs/2013ApJ...769..127L} {769, 127}

\bibitem[\protect\citeauthoryear{{L{\'o}pez-Gonzaga}, {Jaffe}, {Burtscher}, {Tristram}  \& {Meisenheimer}}{{L{\'o}pez-Gonzaga} et~al.}{2014}]{LopezGonzaga2014}
{L{\'o}pez-Gonzaga} N.,  {Jaffe} W.,  {Burtscher} L.,  {Tristram} K.~R.~W.,   {Meisenheimer} K.,  2014, \mn@doi [\aap] {10.1051/0004-6361/201323002}, \href {https://ui.adsabs.harvard.edu/abs/2014A&A...565A..71L} {565, A71}

\bibitem[\protect\citeauthoryear{{Mason}}{{Mason}}{2015}]{Mason2015}
{Mason} R.~E.,  2015, \mn@doi [\planss] {10.1016/j.pss.2015.02.013}, \href {https://ui.adsabs.harvard.edu/abs/2015P&SS..116...97M} {116, 97}

\bibitem[\protect\citeauthoryear{Miller \& Maloney}{Miller \& Maloney}{1963}]{miller1963systematic}
Miller J.~C.,  Maloney C.~J.,  1963, Communications of the ACM, 6, 58

\bibitem[\protect\citeauthoryear{{Nenkova}, {Ivezi{\'c}}  \& {Elitzur}}{{Nenkova} et~al.}{2002}]{nenkova2002}
{Nenkova} M.,  {Ivezi{\'c}} {\v{Z}}.,   {Elitzur} M.,  2002, \mn@doi [\apjl] {10.1086/340857}, \href {https://ui.adsabs.harvard.edu/abs/2002ApJ...570L...9N} {570, L9}

\bibitem[\protect\citeauthoryear{{Nenkova}, {Sirocky}, {Ivezi{\'c}}  \& {Elitzur}}{{Nenkova} et~al.}{2008}]{nenkova2008}
{Nenkova} M.,  {Sirocky} M.~M.,  {Ivezi{\'c}} {\v{Z}}.,   {Elitzur} M.,  2008, \mn@doi [\apj] {10.1086/590482}, \href {https://ui.adsabs.harvard.edu/abs/2008ApJ...685..147N} {685, 147}

\bibitem[\protect\citeauthoryear{{Netzer}}{{Netzer}}{2015}]{Netzer2015}
{Netzer} H.,  2015, \mn@doi [\araa] {10.1146/annurev-astro-082214-122302}, \href {https://ui.adsabs.harvard.edu/abs/2015ARA&A..53..365N} {53, 365}

\bibitem[\protect\citeauthoryear{{Neumayer}, {Cappellari}, {Reunanen}, {Rix}, {van der Werf}, {de Zeeuw}  \& {Davies}}{{Neumayer} et~al.}{2007}]{Neumayer2007}
{Neumayer} N.,  {Cappellari} M.,  {Reunanen} J.,  {Rix} H.~W.,  {van der Werf} P.~P.,  {de Zeeuw} P.~T.,   {Davies} R.~I.,  2007, \mn@doi [\apj] {10.1086/523039}, \href {https://ui.adsabs.harvard.edu/abs/2007ApJ...671.1329N} {671, 1329}

\bibitem[\protect\citeauthoryear{Ram et~al.}{Ram et~al.}{2003}]{ram2003dr}
Ram S.~L.,  et~al., 2003, Dr. Alan Kay on the Meaning of Object-Oriented Programming

\bibitem[\protect\citeauthoryear{{Rejkuba}}{{Rejkuba}}{2004}]{Rejkuba2004}
{Rejkuba} M.,  2004, \mn@doi [\aap] {10.1051/0004-6361:20034031}, \href {https://ui.adsabs.harvard.edu/abs/2004A&A...413..903R} {413, 903}

\bibitem[\protect\citeauthoryear{{Riffel}, {Pastoriza}, {Rodr{\'\i}guez-Ardila}  \& {Bonatto}}{{Riffel} et~al.}{2009}]{Riffel2009}
{Riffel} R.,  {Pastoriza} M.~G.,  {Rodr{\'\i}guez-Ardila} A.,   {Bonatto} C.,  2009, \mn@doi [\mnras] {10.1111/j.1365-2966.2009.15448.x}, \href {https://ui.adsabs.harvard.edu/abs/2009MNRAS.400..273R} {400, 273}

\bibitem[\protect\citeauthoryear{{Riffel} et~al.,}{{Riffel} et~al.}{2022}]{Riffel2022}
{Riffel} R.,  et~al., 2022, \mn@doi [\mnras] {10.1093/mnras/stac740}, \href {https://ui.adsabs.harvard.edu/abs/2022MNRAS.512.3906R} {512, 3906}

\bibitem[\protect\citeauthoryear{{Stalevski}, {Tristram}  \& {Asmus}}{{Stalevski} et~al.}{2019}]{Stalevski2019}
{Stalevski} M.,  {Tristram} K. R.~W.,   {Asmus} D.,  2019, \mn@doi [\mnras] {10.1093/mnras/stz220}, \href {https://ui.adsabs.harvard.edu/abs/2019MNRAS.484.3334S} {484, 3334}

\bibitem[\protect\citeauthoryear{{Thatte}, {Quirrenbach}, {Genzel}, {Maiolino}  \& {Tecza}}{{Thatte} et~al.}{1997}]{Thatte1997}
{Thatte} N.,  {Quirrenbach} A.,  {Genzel} R.,  {Maiolino} R.,   {Tecza} M.,  1997, \mn@doi [\apj] {10.1086/304848}, \href {https://ui.adsabs.harvard.edu/abs/1997ApJ...490..238T} {490, 238}

\bibitem[\protect\citeauthoryear{Van~Rossum, Warsaw  \& Coghlan}{Van~Rossum et~al.}{2001}]{van2001pep}
Van~Rossum G.,  Warsaw B.,   Coghlan N.,  2001, Python. org, 1565

\bibitem[\protect\citeauthoryear{{Wales} \& {Doye}}{{Wales} \& {Doye}}{1997}]{Basinhopping1997}
{Wales} D.~J.,  {Doye} J. P.~K.,  1997, \mn@doi [Journal of Physical Chemistry A] {10.1021/jp970984n}, \href {https://ui.adsabs.harvard.edu/abs/1997JPCA..101.5111W} {101, 5111}

\makeatother
\end{thebibliography}

%%%%%%%%%%%%%%%%%%%%%%%%%%%%%%%%%%%%%%%%%%%%%%%%%%

%%%%%%%%%%%%%%%%% APPENDICES %%%%%%%%%%%%%%%%%%%%%

\appendix

\section{Tests}
\label{appendix}

Here we present three complementary tests to the ones described in Subsec. \ref{subsec:synthetic} using synthetic spectra. In these cases we consider different relative proportions of S/N between the Target and Reference spectra. The construction of the spectra follow the same steps as described before: fixed parameter values of $\alpha = 3.5$, $\beta= 8.3$, and $\gamma = -3.3$, and temperatures values ranging between 500 and 1600\,K. The stellar population of the Reference spectrum is modeled as a decreasing linear function.

\begin{figure*}
    \includegraphics[width=.75\linewidth]{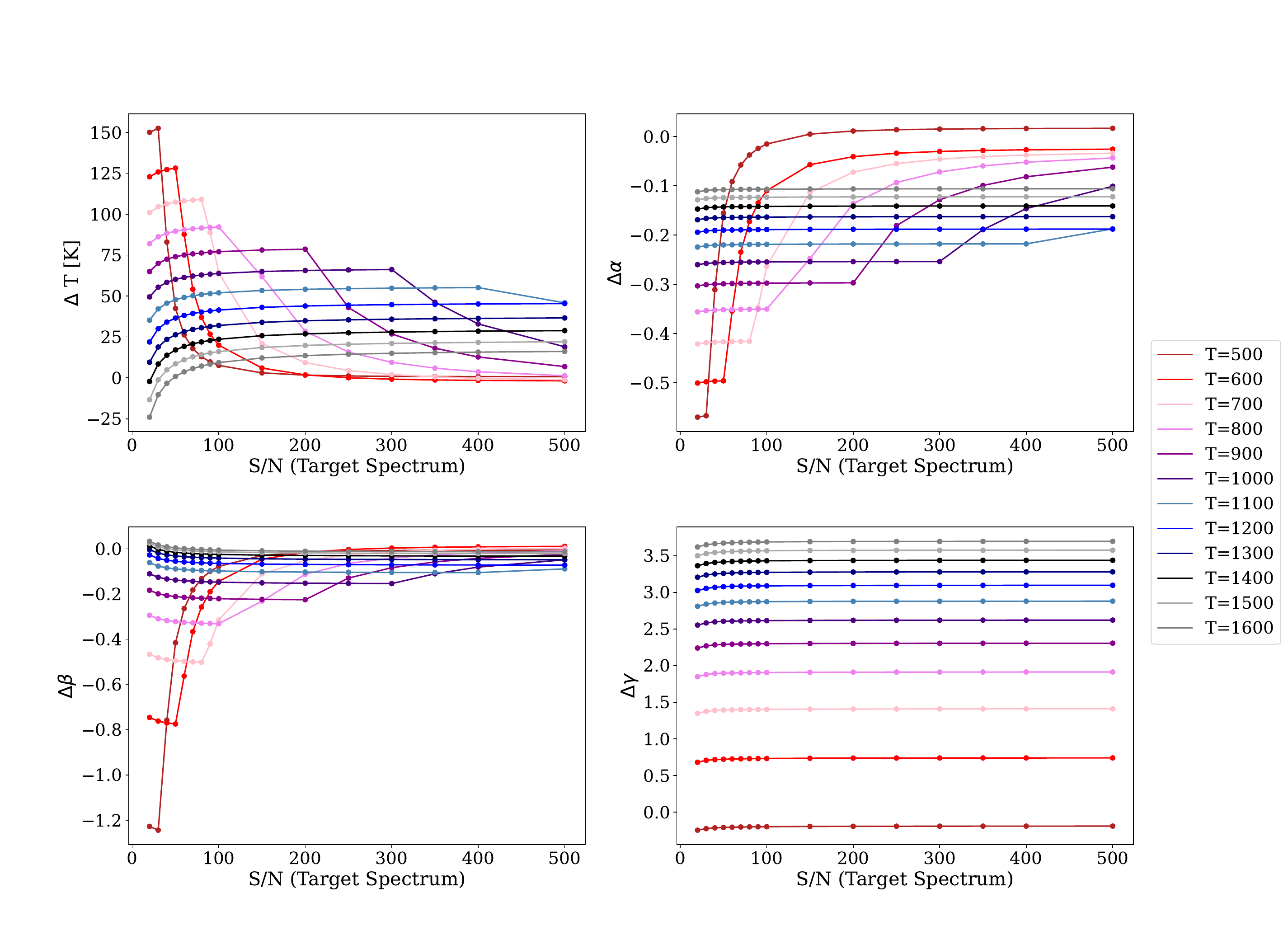}

\caption{Uncertainty in the estimated parameters when varying the S/N of both the Target and the Reference spectra for different temperatures. The temperatures showed in the legend are the temperatures of the synthetic Target Spectrum in each case, the S/N of the Reference Spectrum corresponds to 75\% of the S/N of the Target Spectrum.}
\label{fig:snr75}
\end{figure*}

\begin{figure*}

    \includegraphics[width=.75\linewidth]{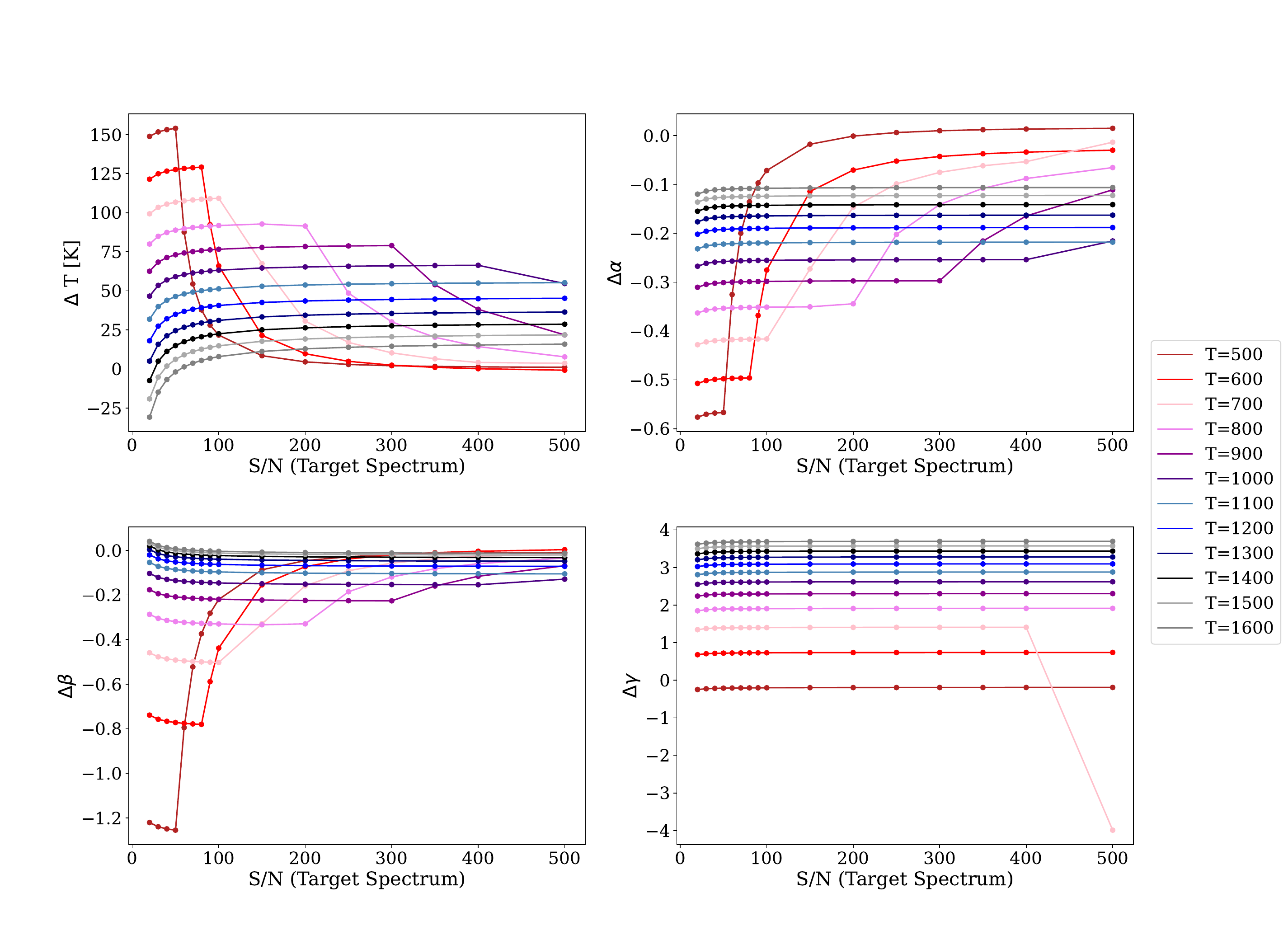}

\caption{ Same as in Fig. \ref{fig:snr75} but the S/N of the Reference Spectrum corresponds to 50\% of the S/N of the Target Spectrum.}
\label{fig:snr50}
\end{figure*}

\begin{figure*}

    \includegraphics[width=.75\linewidth]{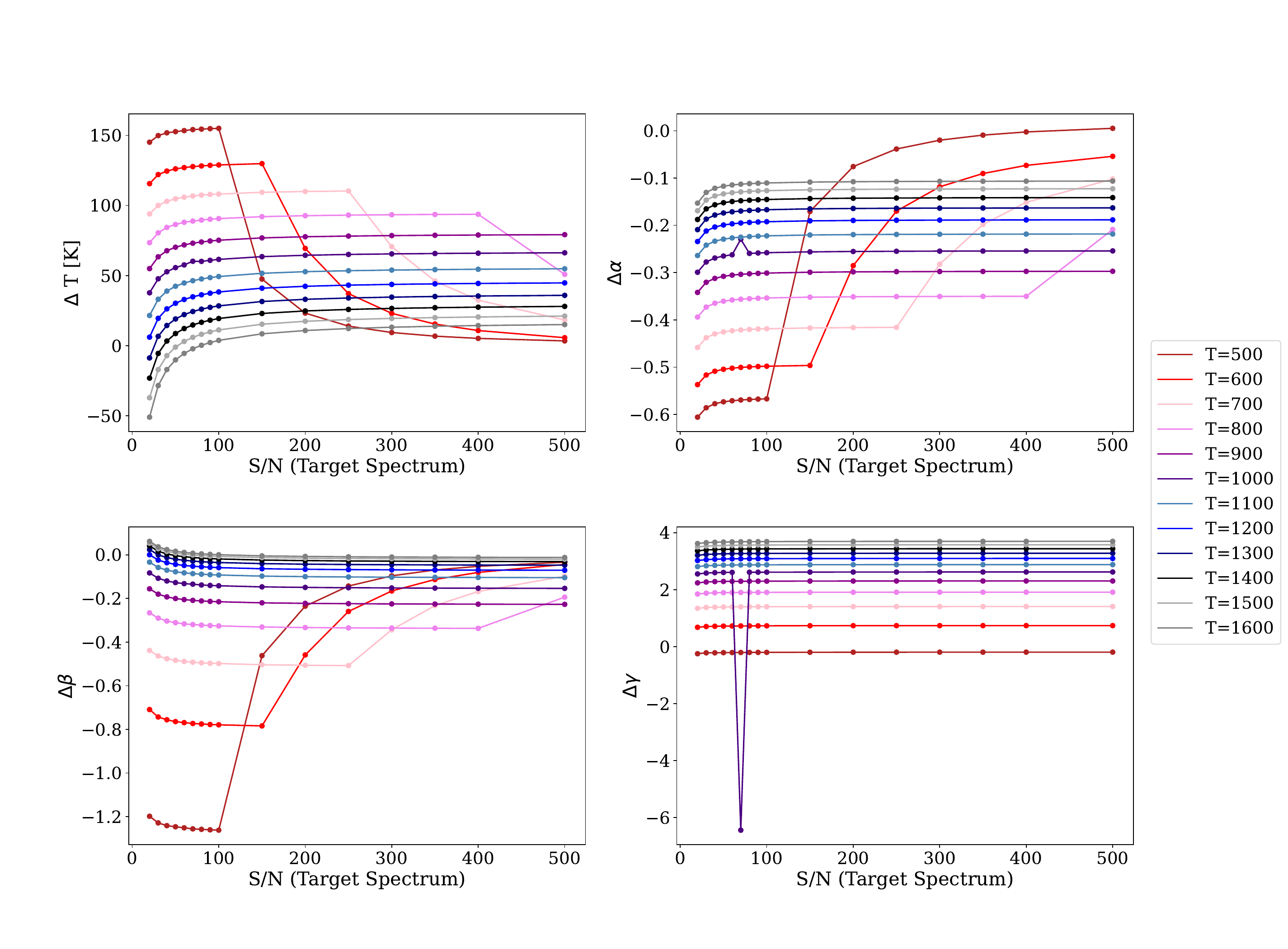}

\caption{Same as in Fig. \ref{fig:snr75} but the S/N of the Reference Spectrum corresponds to 25\% of the S/N of the Target Spectrum.}
\label{fig:snr25}
\end{figure*}

In Figures \ref{fig:snr75}, \ref{fig:snr50} and \ref{fig:snr25} we present the results for the scenarios where the S/N of the Reference Spectrum is $75\%$, $50\%$ and $25\%$ of the S/N of the Target Spectrum, respectively. 

In the three cases the behaviour of the parameters T, $\alpha$, $\beta$, and $\gamma$ is similar to the case where the S/N of both spectra is equal (Fig. \ref{fig:snr}). However, a slight difference is noticeable in the low S/N regime: as the S/N fraction of the Reference Spectrum decreases, the point where the lowest temperatures raise their uncertainty occurs at higher Target Spectrum S/N.

% =======================================================================00

\section{Application example}
\label{Apendix:application_example}

In this subsection we present an application example of the simplest case scenario. A more complete example can be found in the Github repository of the project\footnote{\url{https://github.com/Gaiana/nirdust}}. In this example the fitting is performed over the nuclear spectrum of the galaxy but the reader should keep in mind that NIRDust can be applied at any radius where the dust component is present.

\begin{figure}
    \includegraphics[width=1\linewidth]{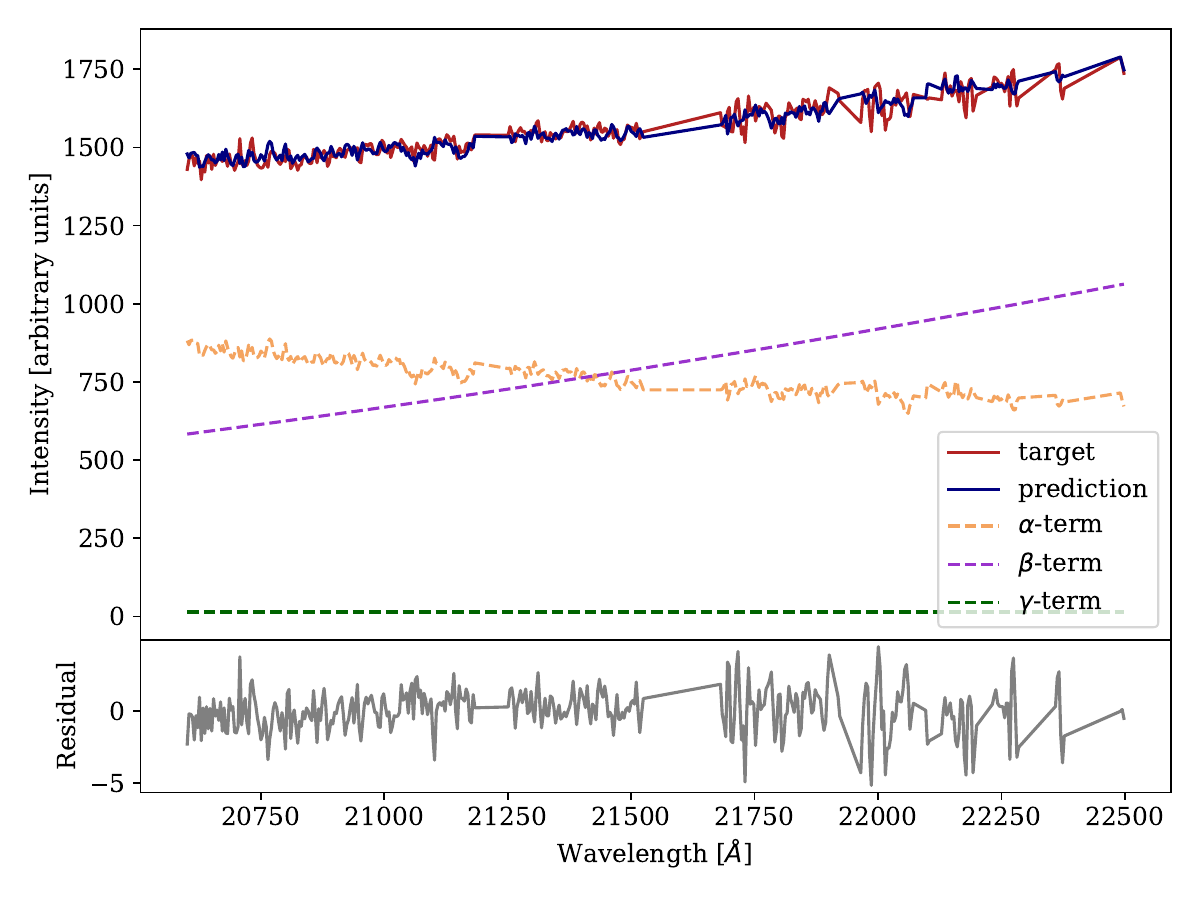}

\caption{Plot showing the result of the fitting procedure as given by the method \texttt{plot()} of the class \texttt{NirdustResults}. The fitting was performed over the Flamingos-2 dataset described in Sec. \ref{subsec:comparison}. The individual components involved in the modeling process of equation \ref{eq2}, labeled as $\alpha$, $\beta$ and $\gamma$ terms, are shown in the top panel. The residual of the fit is shown in the bottom panel and it is consistent with zero. The masked parts in the Target and Reference spectra are easily recognized as straight lines.}
\label{fig:ejemplo}
\end{figure}

The first step is to read the spectrum from a FITS file, providing the name of the file and the redshift of the galaxy. Then it is necessary to follow some steps in order to “clean” the spectrum of undesirable spectral features such as emission/absorption lines or residuals from the reduction process. The user may need to cut the borders of the spectrum due to higher noise or to remove spectral features such as the CO band usually present in the end of the $K_{long}$ filter. All this steps can be performed with the functionalities described in the previous subsection. The impact of these operations on the best-fit solution is evaluated in Section~\ref{sec:performance}. These procedures must be followed for both spectra: the Target and the Reference.
Once both spectra are pre-processed, the function \texttt{fit\_blackbody()} can be executed to obtain the model that best describes the data. The initial values and bounds for the parameters can be provided by the user along with the amount of iterations required. As well, the contribution of the $\gamma$ term can be limited through the \texttt{gamma\_target\_fraction} parameter. In this example the bounds, the amount of iterations, and the limit on the $\gamma$ term are initially given to the function. In Fig.~\ref{fig:ejemplo} an example of the output plot is presented. 

\begin{lstlisting}[language=Python]
>>> import nirdust as nd
>>> import matplotlib.pyplot as plt

# Read the two spectra from the FITS file
# the redshift for this galaxy is 0.00183
>>> target_spectrum = nd.read_fits(
...  "nuclear_spectrum.fits", z=0.00183)

>>> ref_spectrum = nd.read_fits(
...  "external_spectrum.fits", z=0.00183)

# Cut the borders of the spectrum in Angstroms
# This is user defined
>>> cut_tgt = target_spectrum.cut_edges(
...  20600, 22700)

>>> cut_ref = ref_spectrum.cut_edges(
...  20600, 22700)

# Find spectral lines and construct the masks
>>> lines_tgt, intervals_tgt = \
...  nd.line_spectrum(cut_tgt, noise_factor=5.5)

>>> lines_ref, intervals_ref = \
...  nd.line_spectrum(cut_ref, noise_factor=3.5)

# Apply the mask to the spectra
>>> mask_tgt = cut_tgt.mask_spectrum(
...  line_intervals=intervals_tgt)

>>> mask_ref = cut_ref.mask_spectrum(
...  line_intervals=intervals_ref)

# Match the spectral axis of both spectra
# so they are equally sampled:
>>>  tgt, ref = nd.match_spectral_axes(
...    mask_tgt, mask_ref)

# Fit
>>> bounds = (
...  (400, 2000.0),  (  0, 20), 
...  (  6,     20),  (-10, 10))

>>> result = nd.fit_blackbody(
...  tgt, ref, niter=700, 
...  gamma_target_fraction=0.01,
...  bounds=bounds)

# Obtain the value of the fitted temperature
>>> T = result.temperature.value

# Visualize the hot dust spectrum and the model
>>> axis = result.plot(show_components=True)
>>> plt.show()

\end{lstlisting}

% testing =====================================================================
\subsection{Quality}
\label{sec:testing}

Quality is a user-determined characteristic and not a technical feature, which is based on the user's experience with a product with respect to their interests and requirements, which may vary over time \citep{feigenbaum1983total}. 
From developer perspective software quality can be composed of qualitative and quantitative metrics. For NIRDust  we determine our quality threshold by defending a coding standard, code coverage, unit tests and continuous integration.

We chose to follow the PEP-8 standard \citep{van2001pep}, which defines a coding style convention proposed by the Python community, both for the standard language library and in community projects. Following PEP-8 ensures that any Python programmer has a reasonable understanding of the code written and makes it easier to find collaborators and maintain the project over time.
This standard are enforced using the Flake-8 tool \footnote{\url{https://flake8.pycqa.org}}, that automatically  checks deviations in style and will help minimize the code-errors in future versions.

In addition to style and maintainability, it is clearly obvious that we want NIRDust to work properly on all computers on which it is deployed, taking into account the heterogeneity of operating systems and Python versions. For this reason, we have implemented 93 Unit-testing to validate  that  the  individual  components  of  the  software  work  correctly \citep{Jazayeri2007}. NIRDust is tested for Python 3.8, 3.9 and 3.10, reaching  99$\%$  of code-coverage \citep{miller1963systematic}, or in other words our test-suite executes almost all the code of the project.

The source code is hosted in a public GitHub repository \footnote{\url{https://github.com/Gaiana/nirdust}} with a open-source MIT-License \footnote{\url{https://opensource.org/licenses/MIT}}
The committed versions of code are automatically tested with a continuous-integration workflow implemented with GitHub Actions\footnote{\url{https://github.com/Gaiana/nirdust/actions}} \citep{fowler2006continuous}, and the documentation is built from the repository and made public in the read-the-docs service\footnote{\url{https://nirdust.readthedocs.io/}} \citep{holscher2010read}.

The project is available in the Python-Package-Index (PyPi)\footnote{\url{https://pypi.org/project/nirdust/}} throught the pip installer \footnote{\url{https://pypi.org/project/pip/}}. Finally, the project is registered in the Astrophysics Source Code Library \citep[ASCL,][]{Allen2015}\footnote{\url{https://ascl.net/code/v/3419}}.

\color{black}
%%%%%%%%%%%%%%%%%%%%%%%%%%%%%%%%%%%%%%%%%%%%%%%%%%

% Don't change these lines
\bsp	% typesetting comment
\label{lastpage}
\end{document}